\documentclass[%
preprint,
amsmath,amssymb,
aps,
prb,
floatfix,
]{revtex4-2}

\PassOptionsToPackage{backend=biber}{biblatex}
\usepackage[T1]{fontenc}
\usepackage[utf8]{inputenc}

\usepackage{siunitx}
\usepackage{amsmath} 
\usepackage{bm}

\usepackage{xcolor}
\usepackage[pdfencoding=auto]{hyperref}
\hypersetup{colorlinks=true, citecolor=blue, urlcolor=blue, linkcolor=black}

\usepackage{changepage}

\usepackage{graphicx} 
\usepackage{caption}
\usepackage{subfig}
\usepackage{wrapfig}
\usepackage[export]{adjustbox}

\usepackage{relsize}
\usepackage{multirow}
\usepackage{dcolumn}
\usepackage{array}

\usepackage[normalem]{ulem}
\usepackage{booktabs}
\usepackage{makecell}

\newcommand{\subfigref}[2][\figurename~]{#1\ref{#2}}
\newcommand{\subfigrefs}[3][\figurename~]{#1\ref{#2}~and~\ref{#3}}

\newcommand{\figref}[2][\figurename~]{#1\ref{#2}}

\newcommand{\tabref}[2][Table~]{#1\ref{#2}}

\renewcommand{\eqref}[2][Eq.~]{#1(\ref{#2})}

\newcommand{\etal}{\emph{et al.}}

\DeclareSIUnit\angstrom{\protect\text{Å}}

\newcolumntype{P}[1]{>{\centering\arraybackslash}p{#1}}

\definecolor{amethyst}{rgb}{0.6, 0.4, 0.8}
\newcommand{\red}[1]{{\color{red}{#1}}}

\begin{document}

\title{Atomistic Structure Generation and Neural-Network Screening of Hard Carbons to Identify High-Capacity Sodium Storage}

\author{Harry Mclean}

\author{Aiden Daniel Emery}

\author{Theodore Thomas Walton}

\author{Ned Thaddeus Taylor}

\author{Steven Paul Hepplestone}
\email{s.p.hepplestone@exeter.ac.uk}
\thanks{Corresponding author}

\affiliation{Department of Physics, University of Exeter, Stocker Road, Exeter, EX4 4QL, United Kingdom}

\date{August 2026}

\begin{abstract}
Hard carbons are established anodes for lithium-ion batteries and leading candidates for sodium-ion batteries, yet their electrochemical performance is governed by a heterogeneous network of graphitic domains, defects, and nanopores that conventional atomistic methods cannot model at the required length scales. We combine universal machine-learned interatomic potentials with the RAFFLE structure-generation framework to construct 13,096 realistic hard carbon models containing up to 4,378 atoms, matching experimentally measured densities, porosities, and sp$^2$/sp$^3$ bonding fractions. Explicit sodium intercalation of representative structures reproduces the characteristic sloping-to-plateau voltage profiles, revealing that capacity increases with decreasing carbon density and increasing porosity. To screen the full library, we train a lightweight neural-network surrogate that predicts capacity directly from the host using frozen universal-potential descriptors augmented by geometric void features. The surrogate identifies high-capacity candidates exceeding 800~\si{\milli\ampere\hour\per\gram}, which are validated by full intercalation calculations. This scalable framework links hard carbon microstructure to sodium-storage performance and provides atomistic design principles for high-capacity anodes. More broadly, the workflow enables systematic exploration of synthesis-dependent amorphous microstructures, including precursor chemistry, pyrolysis, heteroatom doping, and pore engineering, providing a route toward atomistically informed hard carbon design.
\end{abstract}


\maketitle

\section{Introduction}

The transition to net-zero electricity requires pairing intermittent renewables with scalable electrochemical storage to ensure grid reliability~\cite{Attanayake2024,Maslin2025,Murphy2024,Njema2024}. Among the diverse electrode chemistries proposed for next-generation batteries~\cite{Phogat2025,NourEddine2025}, materials that are abundant, inexpensive, and compatible with existing manufacturing infrastructure offer the most practical route to near-term improvements. Carbon-based anodes are therefore particularly attractive. Of these, the commercial standard carbon anode for lithium-ion batteries is graphite, delivering a theoretical capacity of 372~\si{\milli\ampere\hour\per\gram} via LiC$_6$ formation~\cite{Dong2025, Tarascon2001,PerssonPropertiesLiGraphite2010}.  In contrast, for Na, Na$^+$ intercalation is both thermodynamically and structurally unfavourable~\cite{Yao2024} resulting in graphite storing only $\approx$35~\si{\milli\ampere\hour\per\gram}. Thus alternate carbon electrodes, which are cheaper than graphite, more sustainable to produce, and have increased sodium capacity, are too compelling to ignore.

Hard carbon is non-graphitisable (does not reform graphite under high temperatures, $> 1000$~\si{\kelvin}) and consists of disordered regions, randomly oriented graphitic fragments, defects, and nanopores~\cite{DouHardCarbonReview2019,HarrisPerspectivesGraphiticCarbon2005,LiHeteroatomDoping2017}, providing multiple ion-storage environments (intercalation, surface adsorption, pore filling)~\cite{Gan2025,Tan2023} and a mixture of sp$^2$ and sp$^3$ bonding~\cite{Chu2023}.
This heterogeneity enables lithium capacities of 500--700~\si{\milli\ampere\hour\per\gram}~\cite{INABA2009198} and recent experimental advances have pushed the capacity of hard carbon anodes for sodium-ion batteries to $\approx$580~\si{\milli\ampere\hour\per\gram}, which exceeding that of commercial Li-ion graphite~\cite{Kamiyama2021MgOTemplateSynthesis}.
This breakthrough has intensified interest in developing a fundamental understanding of hard carbon, as the complex, coupled storage mechanisms make development of such materials a vastly complicated configuration space.

Despite decades of research, hard carbon optimisation remains largely empirical. Precursor chemistry, pyrolysis temperature, and heating rate are deeply intertwined: they collectively govern graphitic domain size, defect concentration, pore morphology, and the sp$^3$ fraction in ways that are exceptionally difficult to isolate experimentally~\cite{Bommier2014,Shafiee2024}.
Reports have identified a multistage filling process~\cite{Weaving2020ElucidatingSodiumMechanism,Yu2025EngineeringPseudoGraphitic} that ends in pore filling for higher capacities; thus demonstrating that porosity of these hard carbons is critical to determining capacity. To fully understand this link, atomic-scale modelling is ideal, yet considerable care is needed in constructing realistic atomic-structure models.

First-principles methods such as density functional theory have shown that sodium storage mechanisms in hard carbons are strongly density-dependent~\cite{Front2026}. Yet their computational expense restricts systematic exploration to small, ordered unit cells, leaving large disordered models beyond reach. Molecular dynamics (MD) simulations have partially bridged this gap with studies based upon small sample sizes: reactive force-field studies have revealed the staged adsorption, intercalation, and pore-filling of sodium in large-scale hard carbon models~\cite{Li2022} and have probed the thermal and mechanical evolution of nanoporous carbon frameworks~\cite{Busaidi2026}, while specialized refinement techniques have generated amorphous carbon structures spanning densities from porous graphene networks to diamond-like phases~\cite{Bhattarai2018}. However, the persistent disconnect between computationally accessible time and length scales and those probed by experiment~\cite{Madanchi2024} makes the construction of libraries comprising thousands of distinct hard carbon structures prohibitively expensive. Consequently, simplified structural representations, ranging from graphitic-sheet constructs to fullerene-like models of curved graphene networks~\cite{HarrisPerspectivesGraphiticCarbon2005}, have been widely adopted, but these approaches systematically underestimate the topological disorder, microporosity, and structural heterogeneity that dictate real-world sodium storage behavior~\cite{DouHardCarbonReview2019}. In this context, machine-learned interatomic potentials (MLIPs) offer a compelling alternative, leveraging physics-informed training to achieve near-DFT accuracy at MD-scale cost and thereby enabling large-scale statistical sampling of sodium diffusion modes and quantitative structure--transport relationships in disordered carbon anodes~\cite{Rampal2026}.

In this work, we combine universal machine-learned interatomic potentials with the RAFFLE structure-generation framework~\cite{Taylor2025} to construct 13,096 representative hard carbon models containing between 702 and 4,378 atoms (average 2,002 atoms). The generated structures reproduce experimentally relevant densities, sp$^2$/sp$^3$ fractions, graphitic fragments, and pore volumes. Using these models, we systematically investigate how density, porosity, and pore morphology govern sodium-storage voltage and capacity in carbon pure hard carbons. This approach provides atomistic insight into the structure–property relationships that underpin high-performance hard carbon anodes. Finally, we present a method, using a modified frozen MACE-small model, to build a surrogate model for characterising the capacity of all 13,096 hard carbon structures.

\section{Methodology}
To connect hard carbon microstructure with sodium storage at scale, we employ a three-stage computational workflow: structure generation via active-learning random search, explicit intercalation calculations using machine-learned interatomic potentials, and machine-learning-accelerated screening of the full structural library.

Hard carbons are characterised by three principal structural fingerprints: long-range disorder, porosity, and the   $\textrm{sp}^3~\textrm{fraction} = \frac{N_{\textrm{sp}^3}}{N_{\textrm{sp}^2} + N_{\textrm{sp}^3}}.$
where $N_{\textrm{sp}^3}$ ($N_{\textrm{sp}^2}$) is the number of sp$^3$ (sp$^2$) atoms in the cell. Experimental work can often only identify sp$^2$ and sp$^3$, so our sp$^3$ fraction definition reflects this, but we also present the same figures using $\textrm{sp}^{3}/N_{atom}$ in the supplementary for opacity.

Medium-range graphitic order is quantified by Franzblau shortest-path ring analysis~\cite{Franzblau1991}, and formation energies  are reported relative to bulk graphite.  Similarly voltages are calculated with respect to bulk sodium (see Sec.~S1). Following standard practice, we take total capacity as the amount of intercalated sodium at which the voltage falls below 0 V.

Sampling this phase space requires care to avoid biases that would make the dataset unrepresentative. Hard carbon structures were therefore generated with RAFFLE~\cite{Taylor2025}, an active-learning structure-search tool that learns favourable atomic environments from an evolving database of relaxed structures. RAFFLE combines these learned insights with genetic algorithms and multiple placement strategies to guide atom placement within a host, enabling random structure search for systems of 100--1000s of atoms whilst reducing the need to consider every single permutation. Two classes of initial host were used: (1) a single large graphene sheet with a vacuum gap, and (2) multilayer graphene/graphite flakes distributed at random positions and orientations. In some cases, non-interacting templates (spheres and cylinders) were inserted into the periodic host cell to exclude placement from those regions; removing the templates after generation introduced large artificial pores efficiently. Job permutation was automated with the AGOX package~\cite{Christiansen2022AtomisticGlobalOptimization}, using RAFFLE as the generator (See Sec.~S2).

For high-throughput exploration, the energetic cost of structure generation was evaluated with a $\Delta$-learned CHGNet model, implemented by Pitfield \etal~\cite{Pitfield2025AugmentationUniversalPotentials}.
The correction was fitted to 20 carbon structures evaluated with DFT GGA-PBE (see Supplementary Material). During generation, structures were relaxed with the BFGS optimiser for a maximum of 200 steps, with relaxation terminating early if forces fell below 0.05~\si{\electronvolt\per\angstrom}.
Although this criterion is too loose for any single structure, it is effective for rapidly sampling basins of attraction across a broad phase space. Because RAFFLE's generalised descriptor learns from the energetics and structural motifs of each generated and relaxed structure, it is naturally guided toward energetically important regions, whether low-energy basins or undersampled regions of configuration space.
All energetics were subsequently re-evaluated with the MACE-MPA-0 machine-learned interatomic potential~\cite{Batatia2025FoundationModelAtomistic} with the D3 correction included. Structures selected for further analysis were then converged to forces below 0.02~\si{\electronvolt\per\angstrom} using the FIRE optimiser~\cite{Bitzek2006}; this same force tolerance was applied during sodium intercalation.

Structure generation targeted experimentally accessible ranges: sp$^3$ fractions of 0.25--0.55~\cite{Jing_sp2_sp3_ratio2025} and densities of 1.4--2.2~\si{\gram\per\centi\metre\cubed}~\cite{Tang2023}.
In total, 13,096 carbon structures were produced, with unit cells containing between 702 and 4,378 atoms (average 2,002 atoms).

To estimate the uncertainty introduced by the machine-learned interatomic potential, a Monte Carlo analysis was performed. The energies of the intercalated structures were perturbed randomly using an energy variance of 1~\si{\electronvolt\squared} (a conservative estimate; see below), and 1000 samples were generated for each structure. This yielded a maximum capacity standard deviation of 3.98~\si{\milli\ampere\hour\per\gram}.
For context, the variance of carbon-only structures relative to DFT GGA-PBE + vdW is 0.043~\si{\electronvolt\squared}, while that of sodium--carbon structures is 0.012~\si{\electronvolt\squared}; the adopted value of 1~\si{\electronvolt\squared} is therefore deliberately generous.
Because DFT is prohibitively expensive for the large periodic structures generated here, the MACE-MPA-0 potential was validated against DFT on smaller structures to confirm its reliability for this chemical space~FIG.~S1.

Finally, to calculate the capacity of these 13,096 structures, we develop a neural network (NN) surrogate to predict the sodium capacity directly from the unintercalated hard carbon structure. 
This approach is adopted as explicit intercalation calculations for all structures is computationally prohibitive. Each hard carbon structure is encoded by a frozen MACE-MP-0 small model; per-atom descriptors are pooled by mean, maximum, and standard deviation to give a 768-dimensional fingerprint. 
Principal component analysis compresses this to 24 components, to which five geometric void descriptors (void points, void clusters, total void volume, heuristic capacity, and cell volume) are appended. 
The combined vector is standardised and fed into a compact residual multilayer perceptron with a learnable linear skip from the heuristic capacity that anchors predictions to a physical prior. Three output heads predict capacity mean, log-variance, and capacity standard deviation. 
We calculate the capacity of 64 structures, adopting a methodology focused on searching for empty regions in the structure. For a given sodium concentration, all points in the cell lying farther than 2.18~\si{\angstrom} from any atom were identified as candidate void sites. These sites includes pores, interlayer regions, and vacancies. The energetics of 10 randomly sampled combinations of $N_\textrm{sodium}$ filled voids were evaluated with (50 relaxing fire steps, FIG.~S10), with the lowest energy being selected for the voltage calculation. This procedure was repeated for $N_\textrm{sodium}$ ranging from 1 to the target concentration. The voltage profile was then calculated and capacity, was determined by either when sodiation dropped voltage bellow 0 or was extrapolated linearly using the last 3 voltages. 
This method, while less accurate then the one for structures A-D (benchmarked in supplementary material), allowed for the creation of a larger dataset. 
This method allowed for the creation of a larger dataset and is compared with the standard procedure outlined earlier and is benchmarked  for select cases in the Supplementary materials (Sec.~S10).
The neural network model was then trained on the 64 labelled structures with a composite loss of negative log-likelihood, Huber, and mean-squared-error terms, the model achieves a mean absolute error of 21.3~\si{\milli\ampere\hour\per\gram} and $R^2 = 0.946$ under stratified $k$-fold cross-validation. The surrogate was applied to the full library to identify high-capacity candidates, including structure E, for subsequent explicit intercalation calculations.

\section{Results and Discussion}


\begin{figure*}
    \centering
    \subfloat[]{\includegraphics[height=0.43\linewidth]{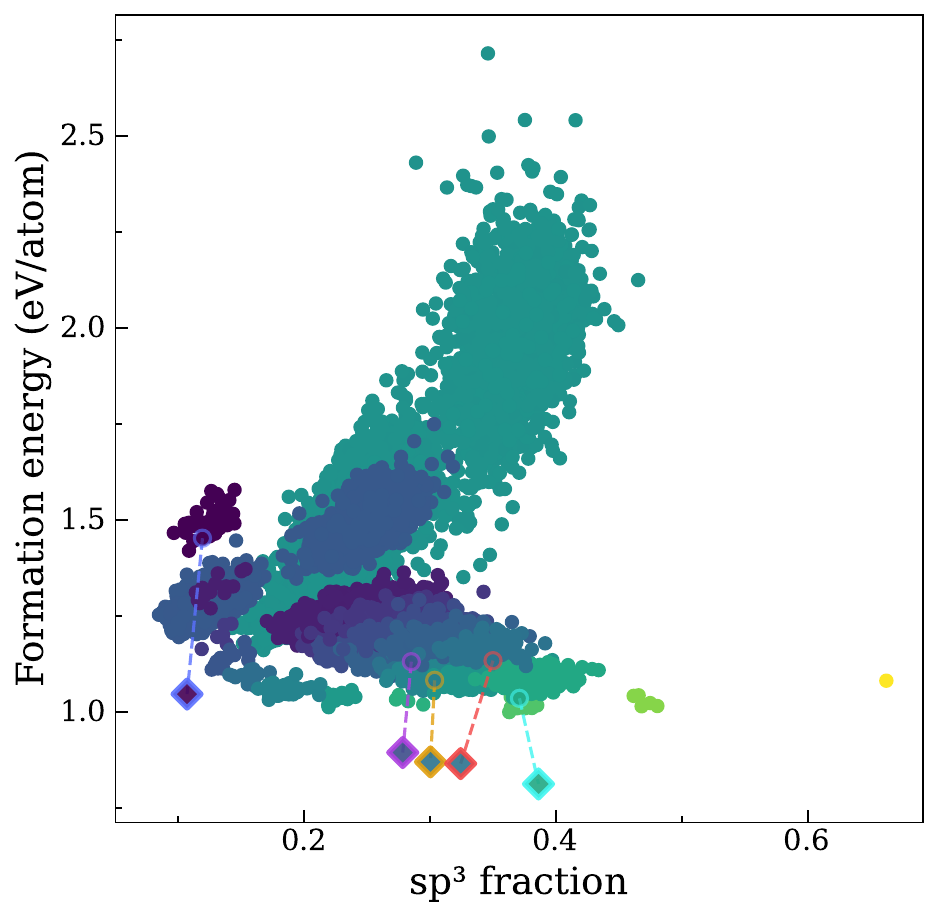}\label{fig:screening:energy}}%
    \hspace{1em}%
    \subfloat[]{\includegraphics[height=0.43\linewidth]{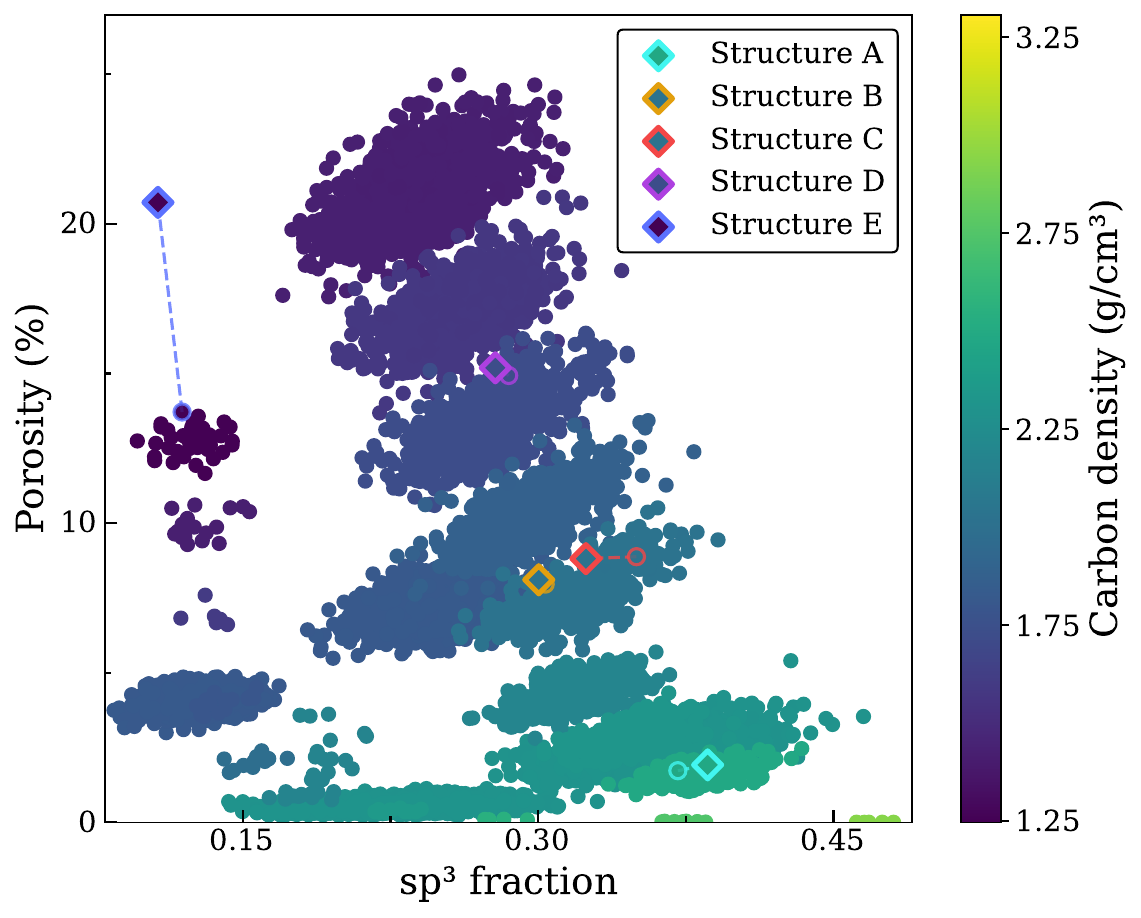}\label{fig:screening:porosity}}
    \hfill
    \subfloat[]{\includegraphics[height=0.43\linewidth]
    {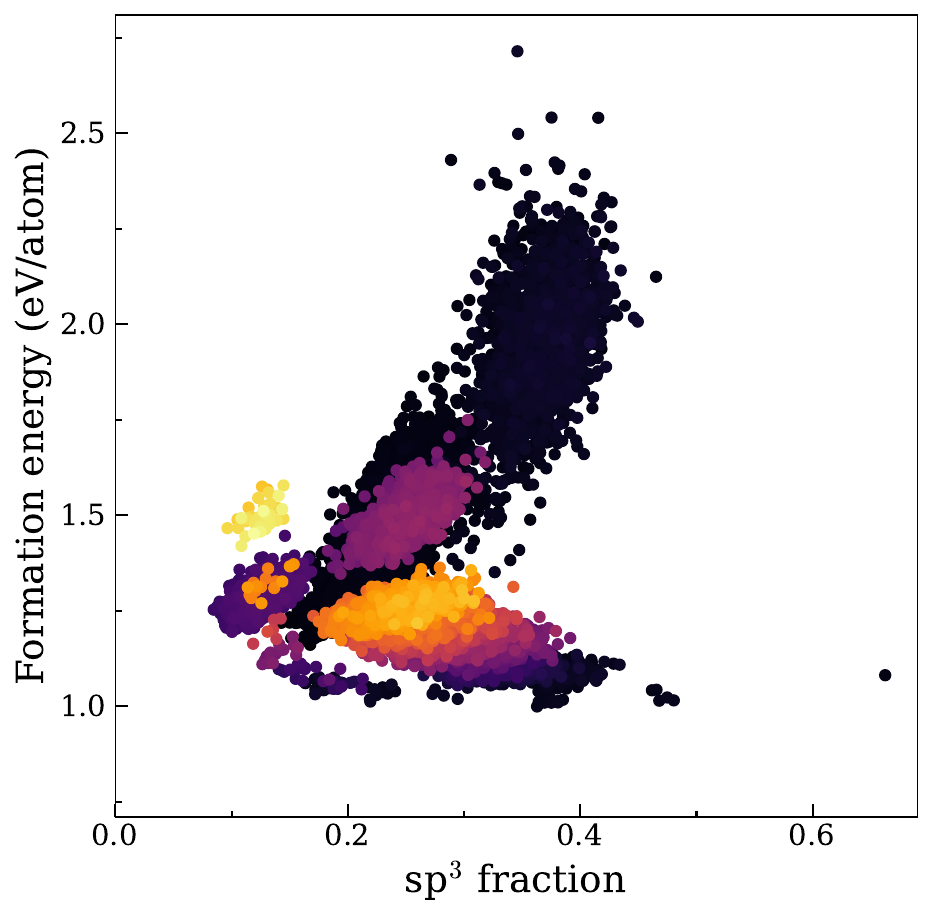}\label{fig:CapacityAll}}%
    \hspace{1em}%
    \subfloat[]{\includegraphics[height=0.43\linewidth]{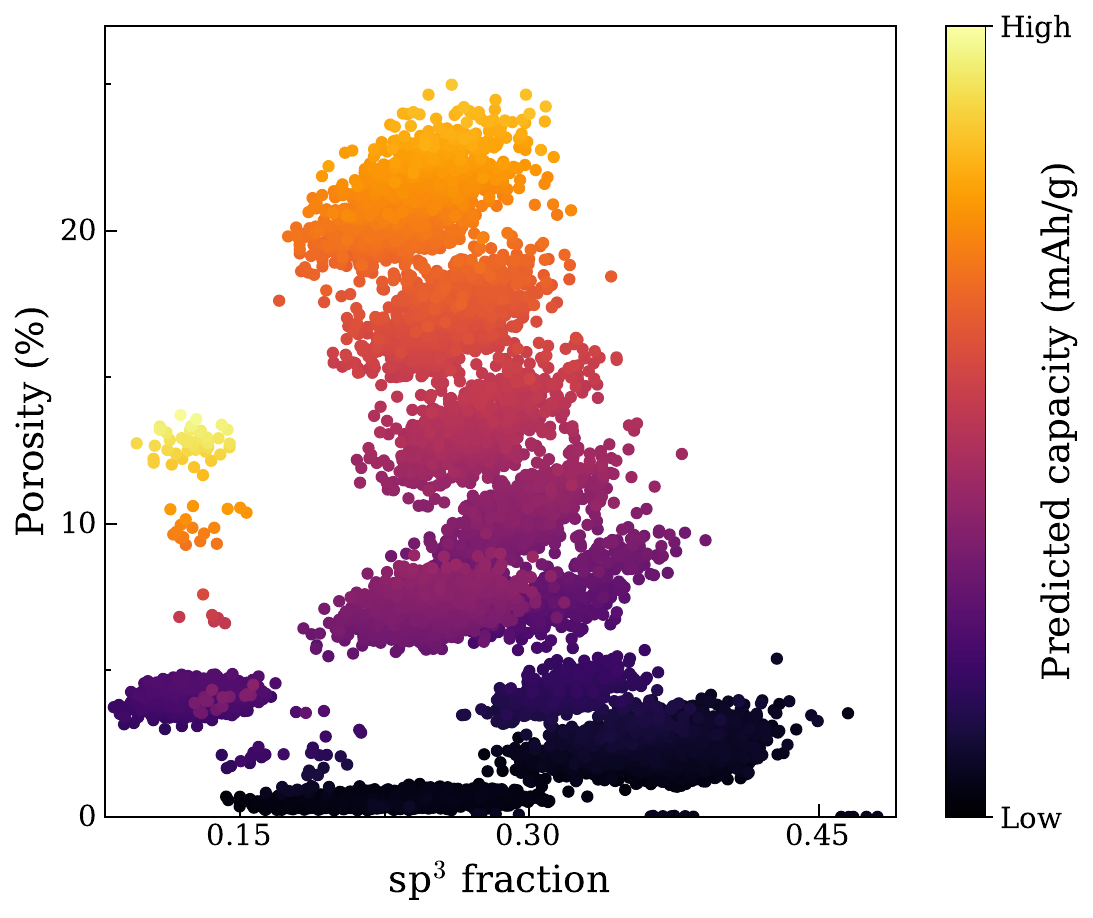}\label{fig:CapacityPorAll}}
    \caption{Structural and energetic analysis of hard carbons generated using AGOX and RAFFLE.
    \protect\subref{fig:screening:energy} Energy per atom (relative to bulk graphite) versus sp$^3$ fraction for 13,096 RAFFLE-generated structures.
    \protect\subref{fig:screening:porosity} Porosity versus sp$^3$ fraction for a subset of 4,926 structures.
    The structures selected for sodium intercalation are highlighted on both plots; the partially (200 BFGS optimizer steps) relaxed (circle) and fully (max force of 0.02~\si{\electronvolt\per\angstrom}) relaxed (diamond) forms for each structure are connected by a dashed line.  Capacity across all 13,096 RAFFLE-generated structures, mapped by energy per atom (relative to bulk graphite, \protect\subref{fig:CapacityAll}) or porosity \protect\subref{fig:CapacityPorAll} versus sp$^3$ fraction using the NN surrogate.}
   
    \label{fig:screening}
\end{figure*}

Experimentally, hard carbons tend to be characterised in terms of the fraction of sp$^2$ (graphite-like) to sp$^3$ (diamond-like) bonding.
This is equivalent to the sp$^3$ fraction which we use to characterise our hard carbon structures (a lower value indicating being more graphite-like).
For the dataset we have created, this fraction is compared to the relative stability, density and porosity of our structures.  
\figref{fig:screening:energy} summarises the energetic and structural properties of all generated structures. 
The hard carbon structures span a broad range of properties: sp$^3$ fractions from 0.085 to 0.662 and densities from 1.246 to 3.307~\si{\gram\per\centi\metre\cubed}, consistent with experimentally explored ranges~\cite{Jing_sp2_sp3_ratio2025,Tang2023}. 
\figref{fig:screening:energy} reveals a wide spread of energies at any given sp$^3$ fraction or density, confirming that the generator samples diverse local bonding environments rather than a single structural family. Comparison of sp$^3$ fraction with porosity (\figref{fig:screening:porosity}) reveals approximately constant-density bands at increasing porosity, confirming that pore volume and density can be tuned largely independently of the sp$^3$ fraction; a weak inverse correlation between porosity and sp$^3$ fraction is apparent.  Also, for the systems generated by our approach, we note that hard carbons with  higher sp$^3$ fraction tend to have low porosity.

For this entire data set, we applied our trained NN to predict the capacity of each structure.  We then selected 5 structures to explore in more detail as a representative set. Our results from the search, shown in \subfigrefs{fig:CapacityAll}{fig:CapacityPorAll} expand the relationships show in \subfigrefs{fig:screening:energy}{fig:screening:porosity}, replacing carbon density with capacity. \subfigrefs{fig:CapacityAll}{fig:CapacityPorAll} reveals that high-capacity regions coincide with low carbon density and high porosity. In  particular, all structures with desnity greater than 2.25~\si{\gram\per\centi\metre\cubed} show very low capacities (lowest 5\% capacity range), whereas structures with density less than 1.5~\si{\gram\per\centi\metre\cubed} possess capacities in the top 33\% of the systems explored.  These results indicate a key threshold for hard carbons with a suitably high working capacity is 2~\si{\gram\per\centi\metre\cubed} or lower.  In particular,  our results in  \subfigref{fig:CapacityPorAll} indicate there exists a "sweet zone" for hard carbons with densities of approximately 1.25~\si{\gram\per\centi\metre\cubed}, sp$^3$ fractions of 0.15 to 0.18 and porosities of 12-15\% which show the highest capacities. 

For exploring how these hard carbon structures perform as sodium electrodes, five hard carbon structures were selected to examine the roles of density, sp$^3$ fraction, and pore morphology. Structures A--E, highlighted in \figref{fig:screening:energy} and \figref{fig:screening:porosity}, were chosen primarily on the basis of differing porosity, and capacity with two deliberate exceptions. Structure C was selected to demonstrate the contrast between a single large pore and a distribution of smaller pores, while structures B and~C share similar porosity but differ in sp$^3$ fraction, enabling these factors to be decoupled. Structures C and D were both prepared using nanotube templating, which leaves a single larger pore upon template removal. Their structural and energetic properties are summarised in \tabref{tab:superpores}. Lastly, Structure E was identified as the highest capacity structure from a surrogate model.


When validated against the explicitly calculated capacities of structures A--E, the surrogate correctly reproduces the ranking ($A < B < C < D < E$) but exhibits quantitative deviations. Structures A--C are under-predicted by 12--46~\si{\milli\ampere\hour\per\gram}, while structures D and E are over-predicted by 42 and 244~\si{\milli\ampere\hour\per\gram}, respectively. 
The large over-prediction for E likely arises because the intercalation protocol does not sample sodium reordering at voltages approaching 0~V, so additional sodium could in principle be accommodated. These errors reflect the limited size and coverage of the training set; nevertheless, the model captures the qualitative trend that capacity increases with porosity and pore connectivity. This machine-learning-accelerated screening confirms pore volume and connectivity as the dominant descriptors of sodium storage, and provides a scalable route to ranking disordered carbon libraries without repeated intercalation calculations.

\begin{figure}
    \centering
    \subfloat[]{\includegraphics[width=0.5\linewidth]{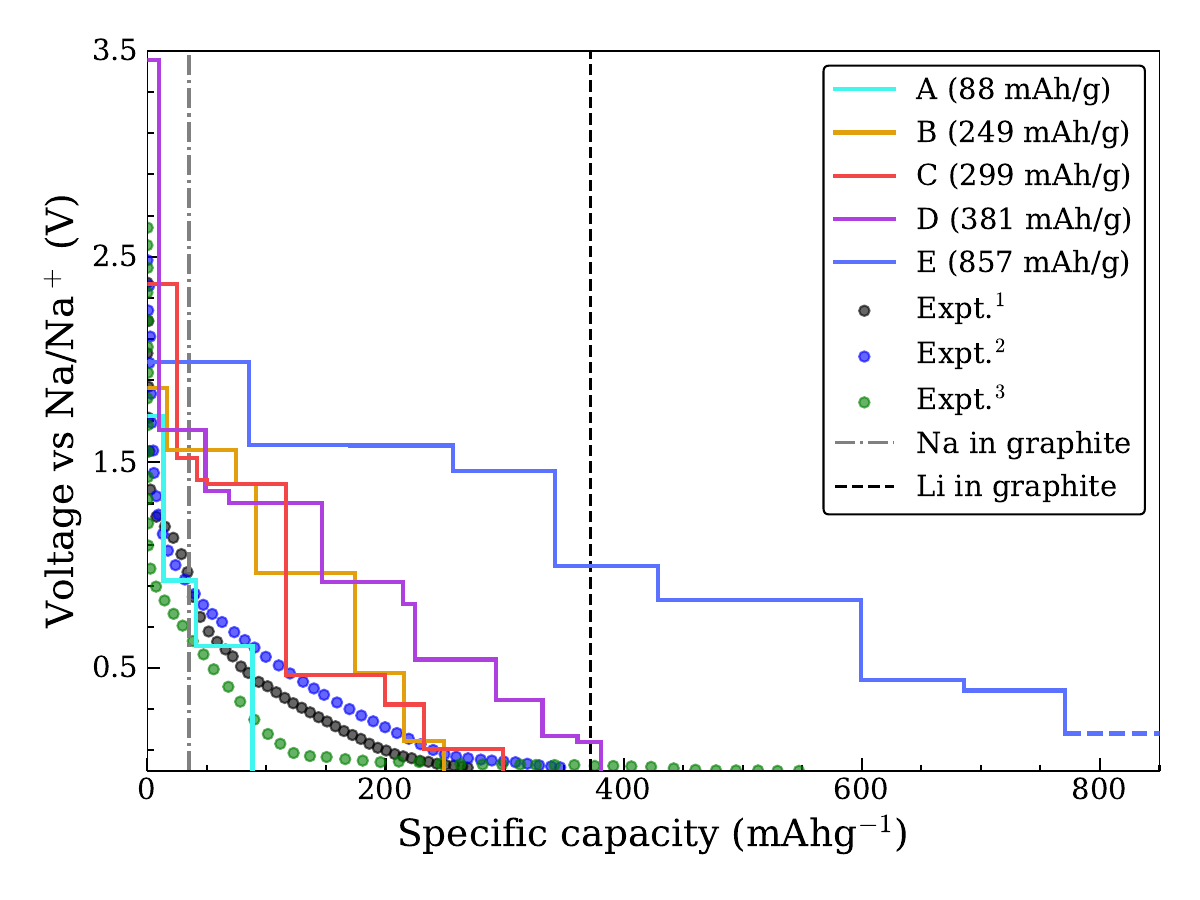}\label{fig:capacity:voltage}}
    \subfloat[]{\includegraphics[width=0.5\linewidth]{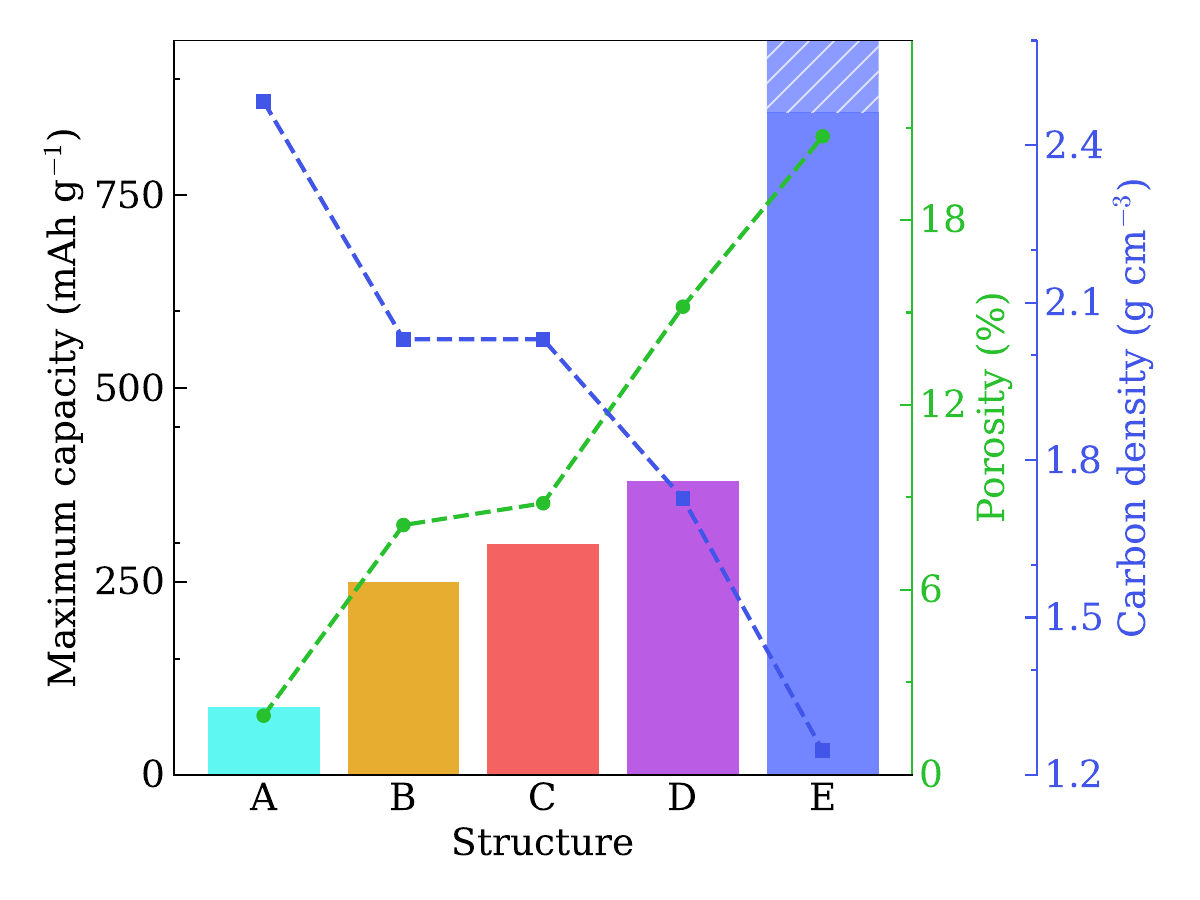}\label{fig:capacity:density}}
    \hfill
    \caption{%
    \protect\subref{fig:capacity:voltage} Computed sodium-filling voltage profiles for structures A--E compared with experimental hard carbon curves (Expt.$^1$, Expt.$^2$, and Expt.$^3$ from \cite{EREN_ExperimentalVoltageCurve2025}, \cite{Zhang2026ThermodynamicKineticInsights}, and \cite{Kamiyama2021MgOTemplateSynthesis}, respectively), sodium intercalation in graphite~\cite{Yao2024}, and lithium intercalation in graphite~\cite{HE2025115426}.
    \protect\subref{fig:capacity:density} Porosity and density versus maximum capacity for structures A--E. The hashed extension of structure E is reflects the non-zero voltage reached.}
    \label{fig:capacity}
\end{figure}

Voltage profiles provide a direct electrochemical test of the generated structures and further validate the NN capacities explored. \figref{fig:capacity} compares the computed sodium-filling curves for structures A--E with experimental hard carbon data~\cite{EREN_ExperimentalVoltageCurve2025,Zhang2026ThermodynamicKineticInsights}, sodium intercalation in graphite~\cite{Yao2024}, and lithium intercalation in graphite~\cite{HE2025115426}.

The experimental hard carbon profile exhibits a high-potential sloping region followed by a low-potential plateau, with maximum capacity approached near 0~\si{\volt}. The simulated structures reproduce this qualitative shape. Structure C tracks the experimental curves below 2.5~\si{\volt} and reaches approximately 300~\si{\milli\ampere\hour\per\gram}, slightly exceeding the experimental value of 275~\si{\milli\ampere\hour\per\gram}~\cite{EREN_ExperimentalVoltageCurve2025}. The maximum capacities of structures A--E are 88, 249, 299, 381, and 857~\si{\milli\ampere\hour\per\gram}, respectively. Structure A has the lowest capacity, consistent with its high density (2.48~\si{\gram\per\centi\metre\cubed}) and small porosity (0.97\%), confirming that sufficient pore volume is essential. Structures B and C share the same density and similar porosities, yet C delivers approximately 50~\si{\milli\ampere\hour\per\gram} additional capacity, demonstrating that pore morphology governs capacity even when density is fixed. Structure E achieves the highest capacity (857.4~\si{\milli\ampere\hour\per\gram}) owing to its high porosity (21\%) and low sp$^3$ fraction (0.11), exceeding both experimental hard carbon references and the capacity of lithium intercalation in graphite~\cite{HE2025115426}.

The variation in the voltage and capacities of these structures is driven by the density and porosity.  Structure A represents the dense extreme (2.48~\si{\gram\per\centi\metre\cubed}), with the highest sp$^3$ fraction (0.386) and the lowest porosity (1.64\%).
Structures B and C have the same density (2.03~\si{\gram\per\centi\metre\cubed}) and nearly identical porosities (8.04\% and 8.79\%), yet they differ in pore architecture: B contains one large pore, whereas C contains a distribution of smaller pores. Their sp$^3$ fractions are also distinct (0.300 versus 0.324), allowing the effects of local disorder and pore topology to be separated.  Structure E sits at the opposite end, with the largest porosity (20.72\%), the lowest density (1.25~\si{\gram\per\centi\metre\cubed}), and a sp$^3$ fraction of 0.107. Despite these structural differences, all five structures share a similar formation energy ($0.86 \pm 0.14$~\si{\electronvolt\per{atom}}). This set therefore provides sufficient separation in density, porosity, and pore topology to test their individual influence on voltage and capacity.

The internal pore architecture of the generated structures was characterised to quantify void connectivity, surface area, and morphology. \tabref{tab:superpores} details the pore properties of structures A--E. The number of pores, pore surface area, and average coordination number reveal clear differences in connectivity. Structure A contains six small, highly interconnected pores (average coordination number 5.0), meaning every pore is connected to all others via tunnels. Structures B and E each contain a single large pore, whereas C and D contain intermediate numbers of larger voids with lower connectivity.

\begin{table}[htbp]
\centering
\caption{Pore properties, graphitic ordering and ring cluster descriptor hard carbon structures A--E. The posority values are given with respect to the unit cell (i.e.\ number of pores per unit cell, surface area per unit cell). The 
graphitic ordering and ring cluster descriptors use  Franzblau analysis of analysis. Bracketed percentage for the largest ring cluster is the percentage of atoms in the unit cell that contribute to said cluster. Defect density $D$, six-ring fraction $f_6$, largest cluster $C_{\text{max}}$ (atoms), and mean cluster size $\langle C \rangle$ (atoms). Clusters are composed of atom rings.}
\label{tab:superpores}

\begin{adjustwidth}{-10cm}{-10cm}
\centering
\begin{tabular}{@{} c c c c c c c c c @{}}
\toprule
Structure & Number & Avg. pore & Surface area per & Avg. coord. & $D$ & $f_6$ & $C_{\text{max}}$ & $\langle C \rangle$ \\
& of pores & volume (\si{\angstrom^3}) & unit cell (\si{\angstrom^2}) & number & & & & \\
\midrule
A & 4 & 53.81 & 1698.85 & 3.00 & 1.27 & 0.33 & 163 (10\%) & 10.42 \\
B & 2 & 552.74 & 5701.15 & 1.00 & 1.19 & 0.34 & 83 (6\%)  & 9.09 \\
C & 6 & 193.45 & 5718.20 & 3.33 & 1.26 & 0.32 & 172 (13\%) & 9.07 \\
D & 3 & 665.68 & 8736.13 & 1.33 & 1.54 & 0.29 & 104 (9\%) & 9.35 \\
E & 1 & 5470.62 & 24237.12 & 0.00 & 1.56 & 0.27 & 35 (2\%) & 7.08 \\
\bottomrule
\end{tabular}
\end{adjustwidth}

\end{table}

The degree of graphitic ordering in each structure was assessed by enumerating the ring-size populations and measuring the connectivity of the resulting graphene-like fragments. \tabref{tab:superpores} summarises the graphitic ordering across the structures. All contain five-, six-, and seven-membered rings, with six-membered rings dominating in A--C and near-equal populations of pentagons and hexagons in D and E. Structure B shows the highest graphitic order ($D = 1.19$, $f_6 = 0.34$), whereas D and E show the poorest ($D \approx 1.55$, $f_6 \approx 0.28$), consistent with their lower densities and greater pore volumes. Despite these differences, mean ring sizes remain near six atoms for all structures. These geometric considerations provide excellent descriptors, which confirm that comparable densities and porosities can yield distinct graphitic orders; the greater ordering of B correlates with a more extended sp$^2$ percolation network than C, despite their similar density and porosity.

\begin{figure*}
    \centering
    \subfloat[Planar density - A]{\includegraphics[width=0.35\linewidth]{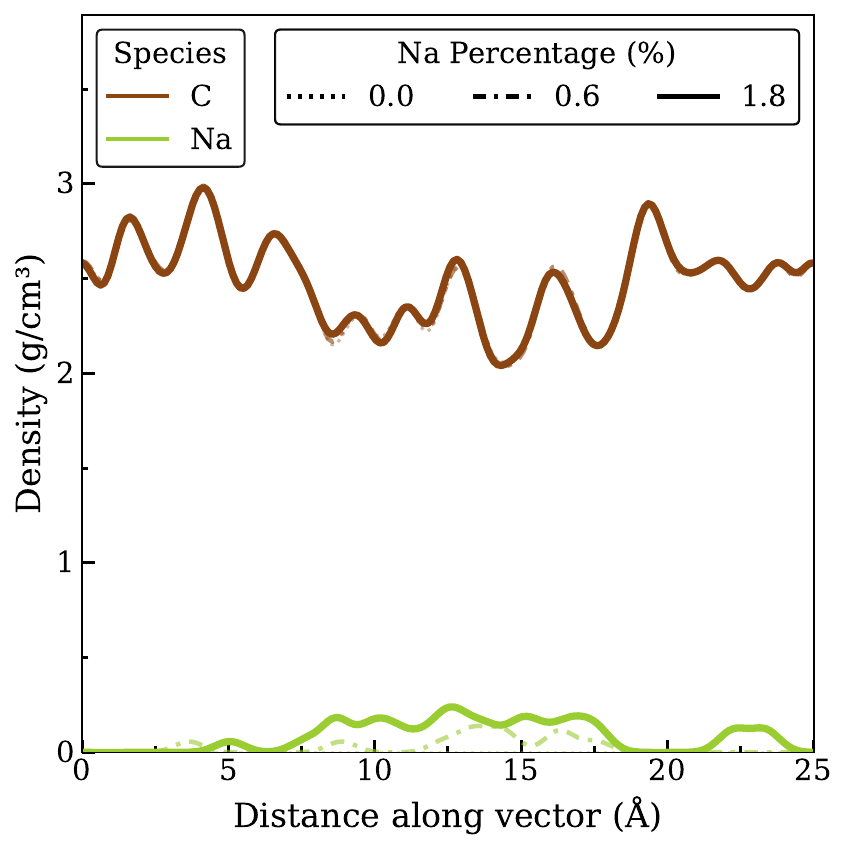}\label{fig:density:A}}\hspace{1em}%
    \subfloat[Pore network - A]{\includegraphics[width=0.47\linewidth]{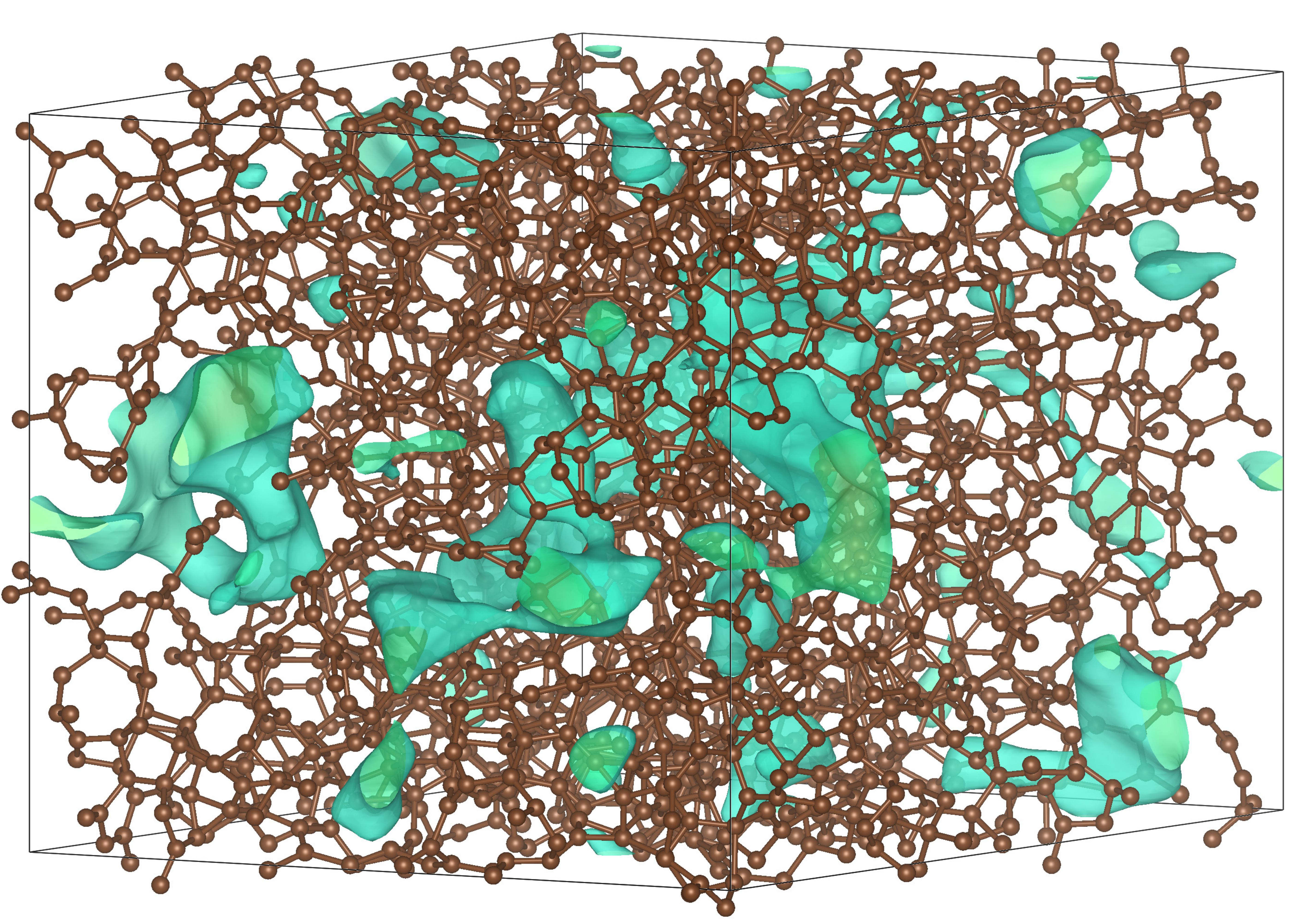}\label{fig:density:A:pores}}
    \hfill{}
    \subfloat[Planar density - D]{\includegraphics[width=0.35\linewidth]{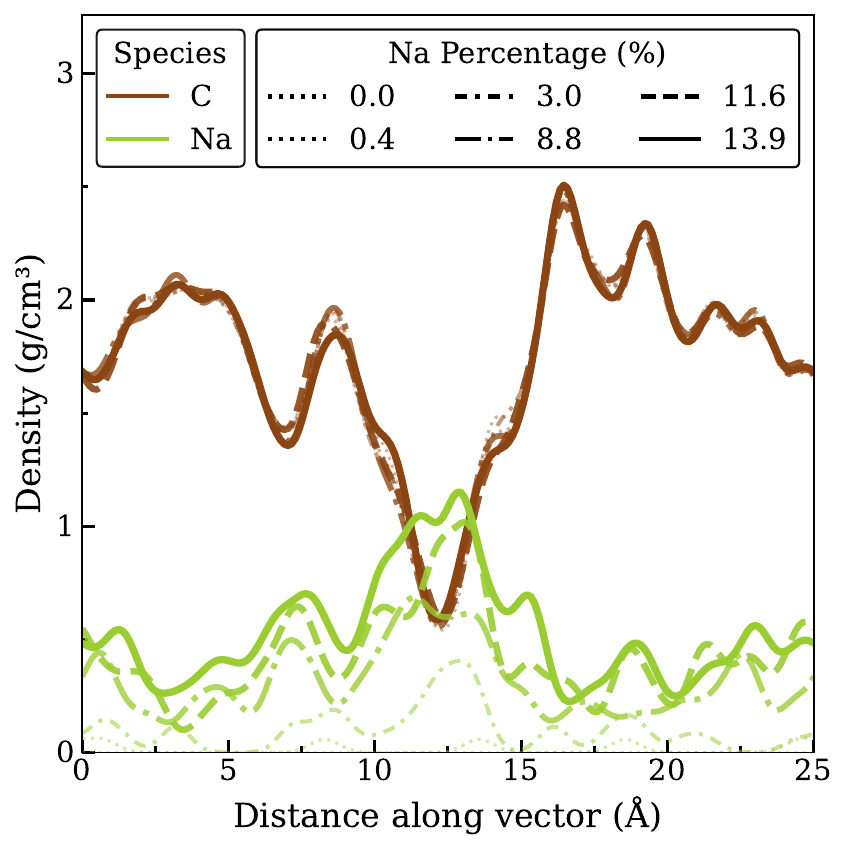}\label{fig:density:D}}\hspace{1em}%
    \subfloat[Pore network - D]{\includegraphics[width=0.47\linewidth]{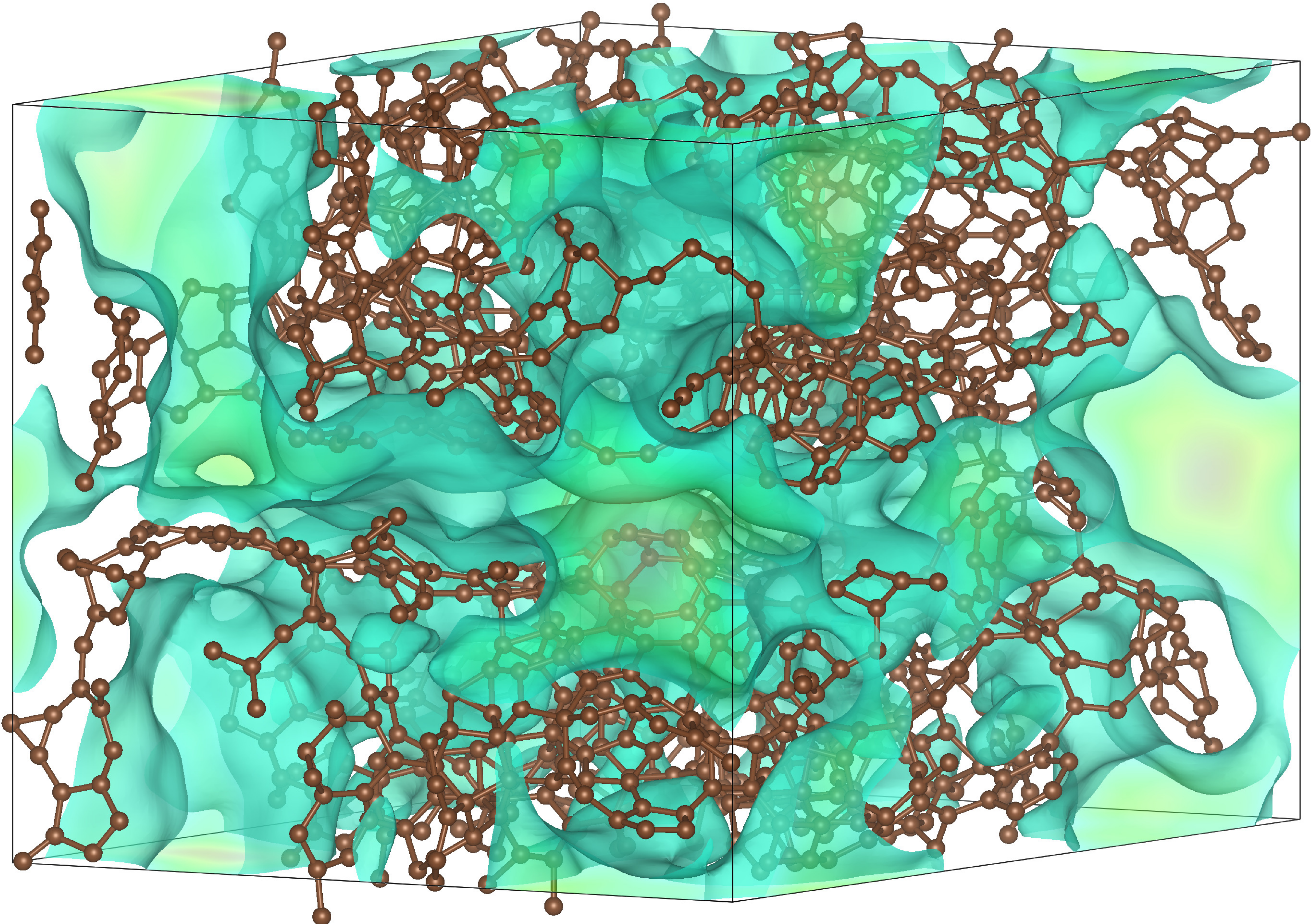}\label{fig:density:D:pores}}
    \caption{%
    Density profile of hard carbon structures \protect\subref{fig:density:A} A and \protect\subref{fig:density:D} D for different levels of sodium intercalation up to their maximum capacities.
    Ball-and-stick model representation of structures \protect\subref{fig:density:A:pores} A and \protect\subref{fig:density:D:pores} D under zero intercalaction with isosurfaces representing the pore network; isosurfaces are rendered for points in space with a minimum atomic distance of roughly 1.8~\si{\angstrom}.%
    }
    \label{fig:density}
\end{figure*}

To understand where sodium is being stored relative to the carbon we provide \figref{fig:density} which presents planar densities (additional results in Supplementary Material, FIG.~S4 and~S5) and pore networks for structures A and D (additional results in Supplementary Material, FIG.~S2), a low and high capacity structure. Structures B,C and E are presented in S2. Structure A exhibits a relatively homogeneous carbon density profile along the $c$-axis under zero intercalation, whereas structure D shows pronounced density oscillations with large void regions. Upon sodiation, sodium accumulates preferentially in regions of low carbon planar density, filling the pore volumes visualised by the isosurfaces. Structure D develops an extensive connected pore network that accommodates substantially more sodium than the isolated micropores of structure A. All structures exhibit increasing carbon reconstruction with higher sodium loading, with the effect becoming more pronounced as porosity increases from A to E. In the dense structure A, intercalation stops before the cell is fully populated, leaving extended regions along the $c$-axis with negligible sodium density.

The spatial distribution of intercalated sodium within each host structure was characterised by planar density analysis at maximum loading. ~Table~S2 summarises the sodium planar density statistics at maximum intercalation. The mean sodium density increases monotonically from structure A to E, reflecting the greater sodium uptake in more porous hosts. The relative standard deviation decreases across the same series, from $76\%$ for structure A to $33\%$ for structure E. This indicates that sodium becomes more uniformly distributed as pore volume increases; in the dense structure A, sodium is confined to sparse isolated sites, whereas in structures D and E the intercalant fills an extensive, connected pore network.

\begin{figure*}
    \centering
    \subfloat[C-C]{\includegraphics[width=0.32\linewidth]{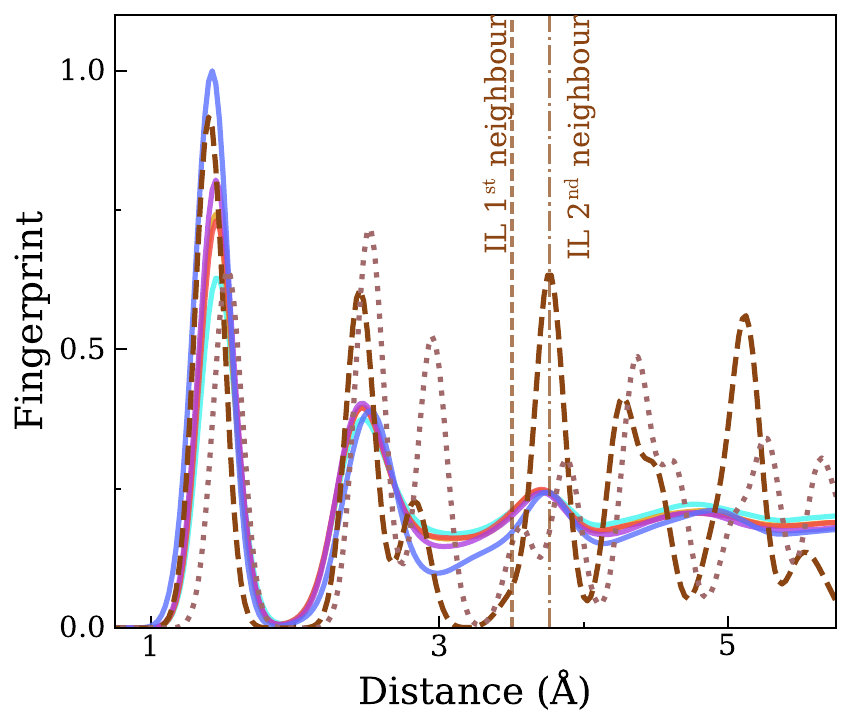}\label{fig:2body:C-C}}
    \subfloat[C-Na]{\includegraphics[width=0.32\linewidth]{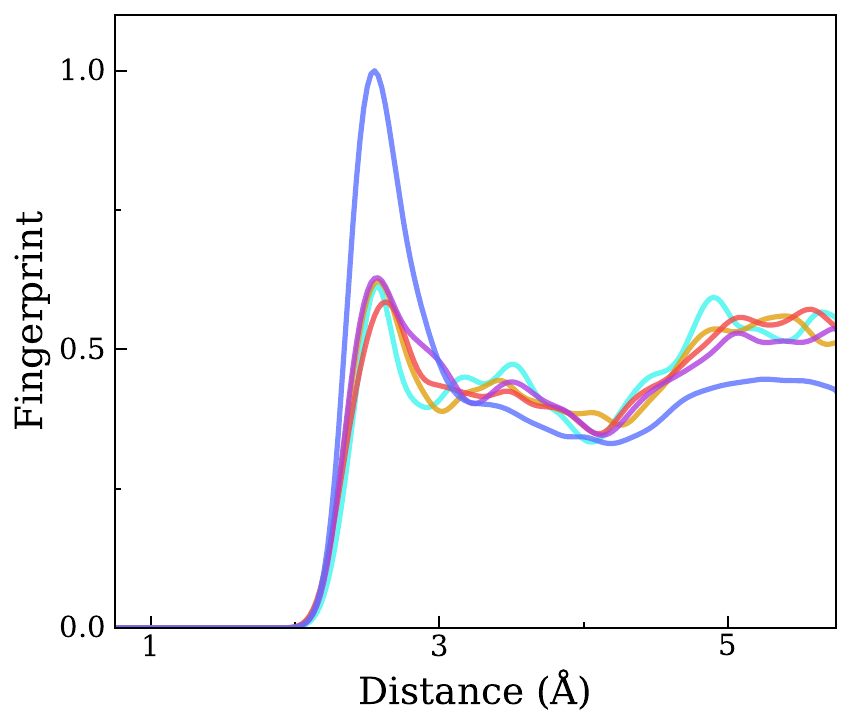}\label{fig:2body:C-Na}}
    \subfloat[Na-Na]{\includegraphics[width=0.32\linewidth]{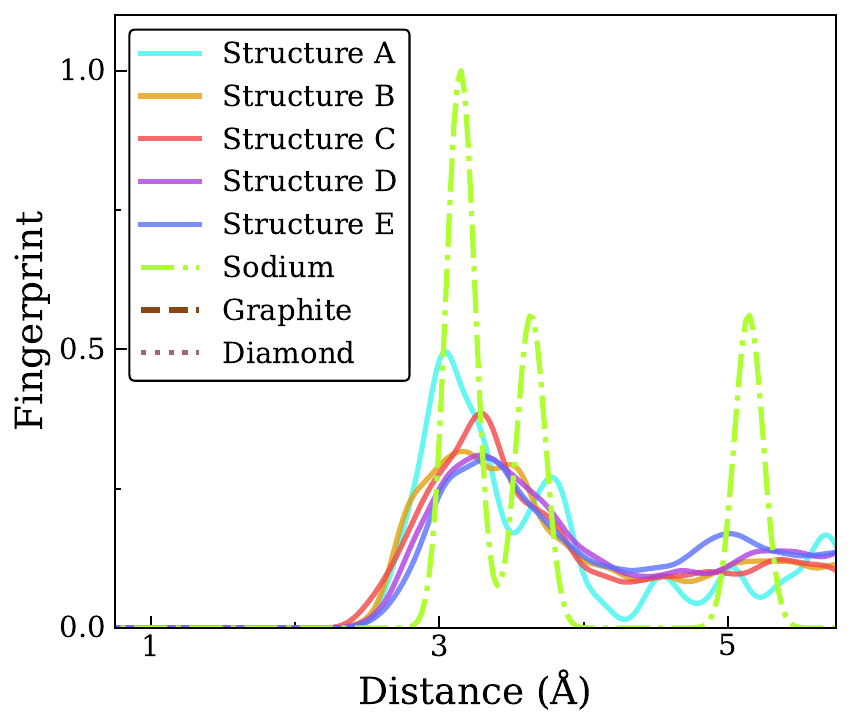}\label{fig:2body:Na-Na}}
    \caption{2-body distribution functions for \protect\subref{fig:2body:C-C} C--C, \protect\subref{fig:2body:C-Na} C--Na, and \protect\subref{fig:2body:Na-Na} Na--Na for structures A--E under their maximum capacity intercalation limit. Reference curves for graphite, diamond, and bulk sodium are included. Fingerprint is normalised to the tallest peak on the plot. IL $n$\textsuperscript{th} neighbour specifies the $n$\textsuperscript{th} inter-layer neighbour bond-length of graphite as obtained from the Materials Project ABC stacked graphite.}
    \label{fig:2body}
\end{figure*}

The short-range atomic order in the generated structures was assessed through two-body distribution functions, which probe local bonding environments and interatomic separations. \figref{fig:2body} presents the RAFFLE 2-body distribution functions for C--C, C--Na, and Na--Na pairs in structures A--E at maximum sodium loading, alongside reference distributions for graphite, diamond, and bulk sodium. All five structures exhibit an identical C--C nearest-neighbour peak at approximately $1.3$~\si{\angstrom}, coincident with graphite and substantially shorter than the diamond value, consistent with a predominantly sp$^2$-bonded network. 
Graphite's interlayer distance of 3.503~\si{\angstrom} (for ABC stacked graphite relaxed using MACE-MPA-0 obtained from the Materials Project~\cite{JainMaterialsProject2013}) is possibly observed in the structures as a broad shoulder of the third peak to the right (at $\approx3.7$~\si{\angstrom}); however, this single peak is heavily suppressed in ABC stacked graphite due to the more dominant 2-body bond of the interlayer next nearest neighbour bond observed at 3.765~\si{\angstrom}. The this also aligns with the interlayer next-nearest-neighbour distance of ABC-stacked graphite relaxed with MACE-MPA-0 ($3.765$~\si{\angstrom}), these results strongly suggesting a graphitic-like stacking is observed locally in the structures despite the structural disorder. The C--Na distributions are nearly identical across all structures and show no systematic trend with host density, indicating that sodium coordinates to carbon in a comparable local environment regardless of porosity.

As for changes in structures, we observe almost identical C--C bonding in all structures, with only differences in the relative magnitude of peaks.
For the fully-intercalated structures, the C--Na bonds show no discernible trend. The Na--Na fingerprints vary markedly across the series. Structure A displays sharp, well-defined peaks at short distances, but this apparent local order is likely an artefact of the small sodium population due to its low sodium site availability, which provides poor statistical averaging. As porosity increases from A to E, the Na--Na nearest-neighbour peak shifts to longer distances and broadens as well as decreases in amplitude. Structure A exhibits a Na--Na distance shorter than that of bulk sodium, whereas structures C and D show distances exceeding the bulk value. None of the structures develop the higher-order peaks characteristic of metallic sodium, confirming that the intercalant remains dispersed rather than forming bulk-like clusters even at maximum loading.

\section{Conclusions}

Our results demonstrate that realistic hard carbon structures with experimentally representative structural and electrochemical characteristics can be generated efficiently using a machine-learning-driven atomistic workflow, and that these models can be analysed at scale to extract design principles for sodium-ion battery anodes. By combining RAFFLE structure generation with machine-learned interatomic potentials, we produce large-scale hard carbon models ($\sim$10$^3$--10$^4$ atoms) spanning experimentally relevant densities, sp$^3$ fractions, graphitic ordering, and pore architectures.

The calculated voltage profiles reproduce the characteristic sloping-to-plateau behaviour across a capacity range of 88--857~\si{\milli\ampere\hour\per\gram}.
Capacity is lowest in the dense, low-porosity structure and highest in the low-density, highly porous one.
The findings agree with the previously reported dependence of capacity on porosity~\cite{Weaving2020ElucidatingSodiumMechanism,Yu2025EngineeringPseudoGraphitic}, where sodium storage begins with intercalation in the sloping region and transitions to pore filling in the plateau region.
The high capacities of Structure D and E (381 \& 857~\si{\milli\ampere\hour\per\gram}, respectively) align well with experimental results of $\sim$580~\si{\milli\ampere\hour\per\gram} (478~\si{\milli\ampere\hour\per\gram} reversible)~\cite{Kamiyama2021MgOTemplateSynthesis} and theoretical predictions of 787~\si{\milli\ampere\hour\per\gram}~\cite{Sun2023}.
Together, these results support the view that connectivity and geometry of internal pore networks should be considered alongside conventional properties such as density, interlayer spacing, and degree of graphitisation~\cite{Front2026,Busaidi2026} when optimising hard carbon electrode performance.

The key methodological advance is a neural network surrogate that predicts sodium capacity directly from the empty host, trained on only 64 structures and achieving excellent accuracy under cross-validation. The model ranks over 13,000 generated structures and identifies high-capacity candidates beyond the initial training set, reducing months of explicit intercalation calculations to seconds per structure. Its success confirms that capacity is governed primarily by pore geometry, and the approach is transferable to other intercalation systems.

Our results show  there exists a "sweet zone" for hard carbons with densities of approximately 1.25~\si{\gram\per\centi\metre\cubed}, sp$^3$ fractions of 0.15 to 0.18 and porosities of 12-15\% which show the highest capacities.   This sweet zone requires not only sufficient pore volume, but requires pores which have sufficient connectivity and elongation. The framework provides a scalable route for systematically linking synthesis parameters to electrochemical performance, moving hard carbon optimisation from empirical trial towards atomistically informed, machine-learning-accelerated materials design.

\section{Data availability}

Data corresponding to this research, including modelling input files and results, will be made available via \red{figshare} upon publication.
The code implementations will be made publicly available under the GNU General Public License version 3 (GPLv3). Scripts and notebooks used to analyse data is made available via a GitHub repository (\url{https://github.com/ExeQuantCode/HardCarbons/}).

\section{Declaration of generative AI and AI-assisted technologies in the manuscript preparation process}

During the preparation of this work the author(s) used Github Copilot in order to improve code documentation and coding efficiency. After using this tool/service, the author(s) reviewed and edited the content as needed and take(s) full responsibility for the content of the published article.

\acknowledgements

We thank the EPSRC for funding H.~Mclean (EPSRC-690010152) and S.~P.~Hepplestone (EP/X013375/1).
N.~T.~Taylor was supported by the Government Office for Science and the Royal Academy of Engineering under the UK Intelligence Community Postdoctoral Research Fellowships scheme (Grant No.~ICRF2425-8-148).

Via our membership of the UK's HEC Materials Chemistry Consortium, funded by EPSRC (EP/R029431), this work used the ARCHER2 UK National Supercomputing Service~\cite{archer} within the framework of a Grand Challenge project.

The authors acknowledge the use of the University of Exeter High-Performance Computing (HPC) facility, ISCA, and resources provided by the Isambard~3 Tier-2 HPC Facility.
Isambard~3 is hosted by the University of Bristol and operated by the GW4 Alliance (\url{https://gw4.ac.uk}), funded by UK Research and Innovation and the Engineering and Physical Sciences Research Council (EP/X039137/1).

\section*{CRediT author statement}

H. M. contributed to the investigation, analysis, software, writing-original draft, writing-review \& editing, data analysis, and visualisation.
A. D. E. contributed to the investigation, analysis, data curation, software, writing-original draft, writing-review \& editing, data analysis, and visualisation.
T. T. W. contributed to the investigation, analysis, data curation, software, writing-original draft, writing-review \& editing, data analysis, and visualisation.
N. T. T. served as project supervisor, providing analysis, data curation, software, writing-original draft, writing-review \& editing, visualisation, funding acquisiton, and project management.
S. P. H. served as project supervisor, providing analysis, writing-review \& editing, and project management.

\bibliographystyle{unsrt}
\bibliography{references}

\end{document}


\title{Supplementary Material: Atomistic Structure Generation and Neural-Network Screening of Hard Carbons to Identify High-Capacity Sodium Storage} 

\author{Harry Mclean}
\author{Aiden Daniel Emery}
\author{Theodore Thomas Walton}
\author{Ned Thaddeus Taylor}
\author{Steven Paul Hepplestone}
\affiliation{Department of Physics, University of Exeter, Stocker Road, Exeter, EX4 4QL, United Kingdom}
\date{August 2026}

\maketitle
\tableofcontents
\clearpage

\section{Energetics and geometry analysis}
\label{sec:energetics}

Medium-range graphitic order is quantified by Franzblau shortest-path ring analysis~\cite{Franzblau1991}, and formation energies are reported relative to bulk graphite,
\begin{equation}
    E_{\textrm{form},i} = \frac{E_{i}}{N_{i}} - \frac{E_\textrm{graphite}}{N_\textrm{graphite}},
    \label{eq:eform}
\end{equation}

where $E_{i}$ and $N_{i}$ are the total energy and atom count of system $i$. Formation energies are reported per atom to enable direct comparison across systems of different size. ABC-stacked graphite is used as the reference because it is the lowest-energy experimentally observed carbon phase, see Materials Project~\cite{JainMaterialsProject2013}. Similarly, the voltages are calculated using 

\begin{equation}
    V = -\frac{E(\mathrm{Na}_{x_2}\mathrm{C}) - E(\mathrm{Na}_{x_1}\mathrm{C}) - (\Delta x)\,E(\mathrm{Na})}{(\Delta x)\,e},
    \label{eq:voltage}
\end{equation}

\noindent
where $E(\mathrm{Na}_{x_2}\mathrm{C})$ and $E(\mathrm{Na}_{x_1}\mathrm{C})$ are the total energies of the hard carbon structures at sodium loadings $x_2$ and $x_1$, respectively, $E(\mathrm{Na})$ is the DFT energy per sodium atom in bulk bcc sodium (the metallic reference), $\Delta x$ is the change in the number of sodium from $E(\mathrm{Na}_{x_1}\mathrm{C})$ to $E(\mathrm{Na}_{x_1}\mathrm{C})$, and $e$ is the elementary charge. ~\eqref{eq:voltage} yields the incremental (stepwise) voltage for the insertion of $\Delta x = x_2-x_1$ sodium equivalents, reported with respect to the Na/Na$^+$ redox couple.

The degree of graphitic ordering in each structure was assessed by enumerating the ring-size populations and measuring the connectivity of the resulting graphene-like fragments. Defect density and six-ring fraction are defined as

\begin{equation}
    D = \frac{N_5 + N_7}{N_6}, \qquad f_6 = \frac{N_6}{\sum_n N_n},
\end{equation}
where $N_n$ is the number of $n$-membered rings. The largest ring cluster is determined using the method of Jain {\it{et al.}}~\cite{Jain2006RingConnectivityMeasuring}. 

\section{RAFFLE calculation details}
\label{sec:raffle}

Multiple permutations of the RAFFLE generator weights were explored. The lowest formation energies were obtained by pre-seeding the host with graphitic fragments (typically of up to 10 fragments of 6-64 atoms with random orientations)  and prioritising the \textit{grow} and \textit{walk} placement methods. In the \textit{walk} method, a trial site is selected at random and nearby points are sampled iteratively until a viable location is found, allowing the search to escape local traps. The \textit{growth} method is a variant that seeds each new search from the previously placed atom, mimicking local nucleation. The RAFFLE search employed an empirical weighting temperature of $k_\mathrm{B}T = 0.2$~eV, a grid spacing of 0.5~\AA\ for viability evaluation, and Gaussian descriptor widths of 0.04~\AA\ (2-body) and $\pi/160$~rad (3- and 4-body). The generator was initialised with transfer data from relaxed graphitic seed structures to accelerate convergence toward realistic carbon bonding motifs.

\section{Validation of the Machine-Learned Potential}
\label{sec:mlip_validation}

\begin{figure}
    \centering
    \includegraphics[width=0.5\linewidth]{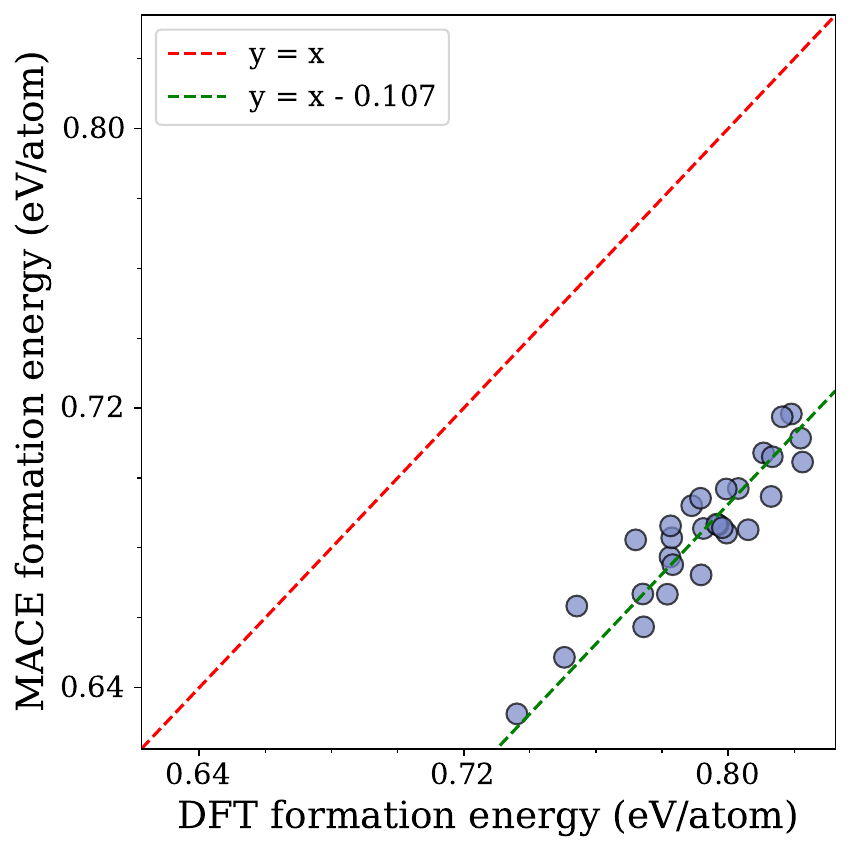}
    \caption{%
    MACE-MPA-0 energies compared with DFT energies for a test set of 30 600-atom hard carbon structures.
    The black red line denotes perfect agreement, and the green dashed line is a linear fit with a root mean squared error of $0.0083$~\si{\electronvolt}.}
    \label{fig:MaceVsDFT}
\end{figure}

\figref{fig:MaceVsDFT} compares MACE-MPA-0 energies with DFT energies for a test set of 600-atom hard carbon structures.
A linear fit to the parity data gives an RMSE of $0.0083$~\si{\electronvolt} and a systematic offset of $-0.11$~\si{\electronvolt}.
The small scatter and approximately systematic bias indicate that MACE-MPA-0 captures relative energetic trends across the generated structures, while the remaining offset could be corrected with a $\Delta$-learning model if higher absolute accuracy were required.

\section{$\Delta$-model}
\label{sec:delta_model}

The $\Delta$-model is a Gaussian process regression model that adjusts the CHGNet energy values to better align with the training dataset.
Here, the training dataset is a set of 20, with 6 additional test, carbon structures with and without defects.
The structures are graphite and graphene structures, nanotubes and buckyballs.
Defects include lithium intercalants and boron and nitrogen substitutions.

The dataset is calculated using DFT GGA-PBE with van der Waals DFT-D3 correction.

The $\Delta$-model improves the CHGNet model MAE from $0.1368$~\si{\electronvolt\per{atom}} to $0.0083$~\si{\electronvolt\per{atom}} on the combined training and test dataset.

\section{Structures}
\label{sec:structures}

\begin{figure*}
    \centering
    \subfloat[Structure B]{\includegraphics[width=0.45\linewidth]{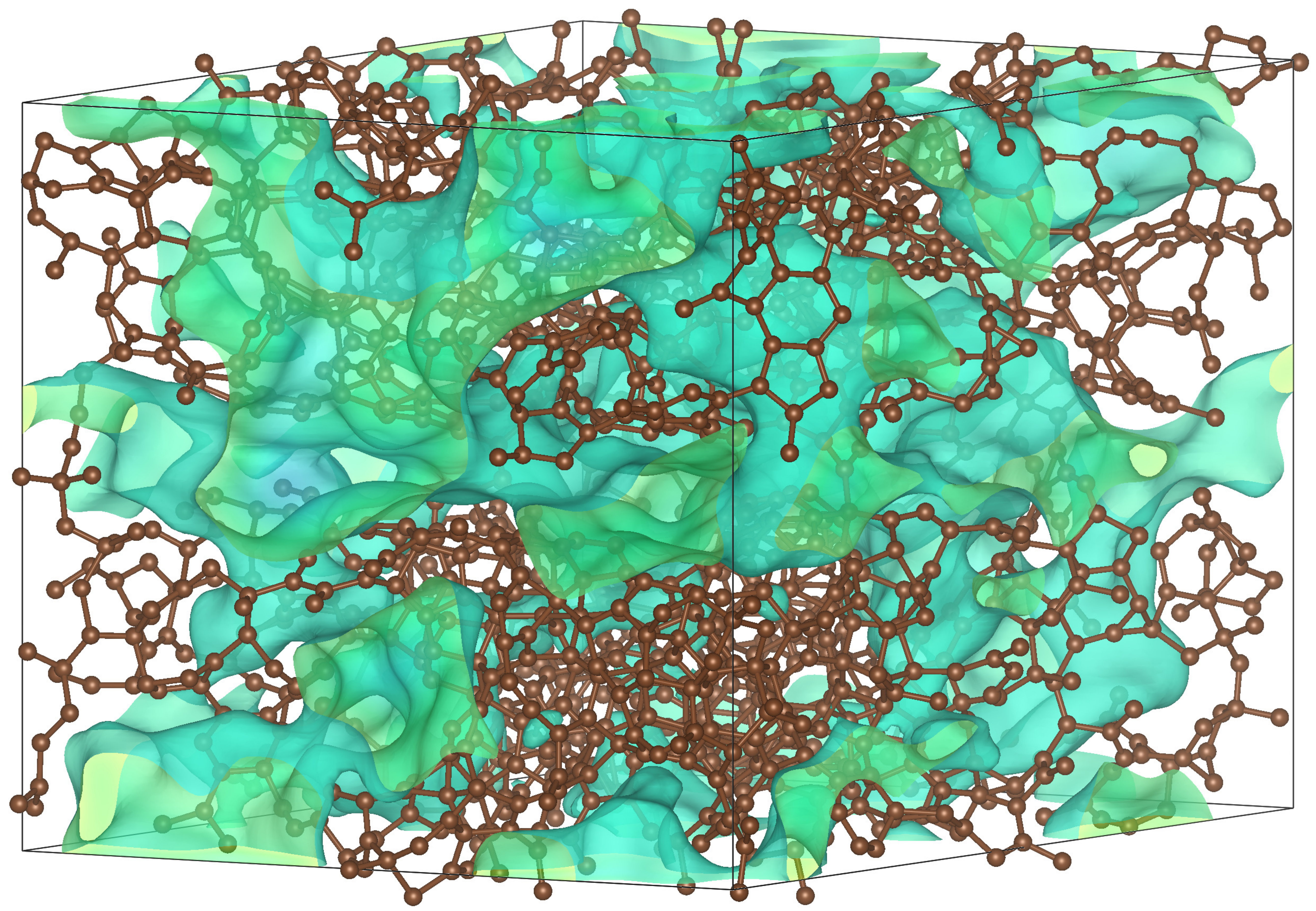}\label{fig:structure:B}}%
    \hspace{1em}%
    \subfloat[Structure C]{\includegraphics[width=0.45\linewidth]{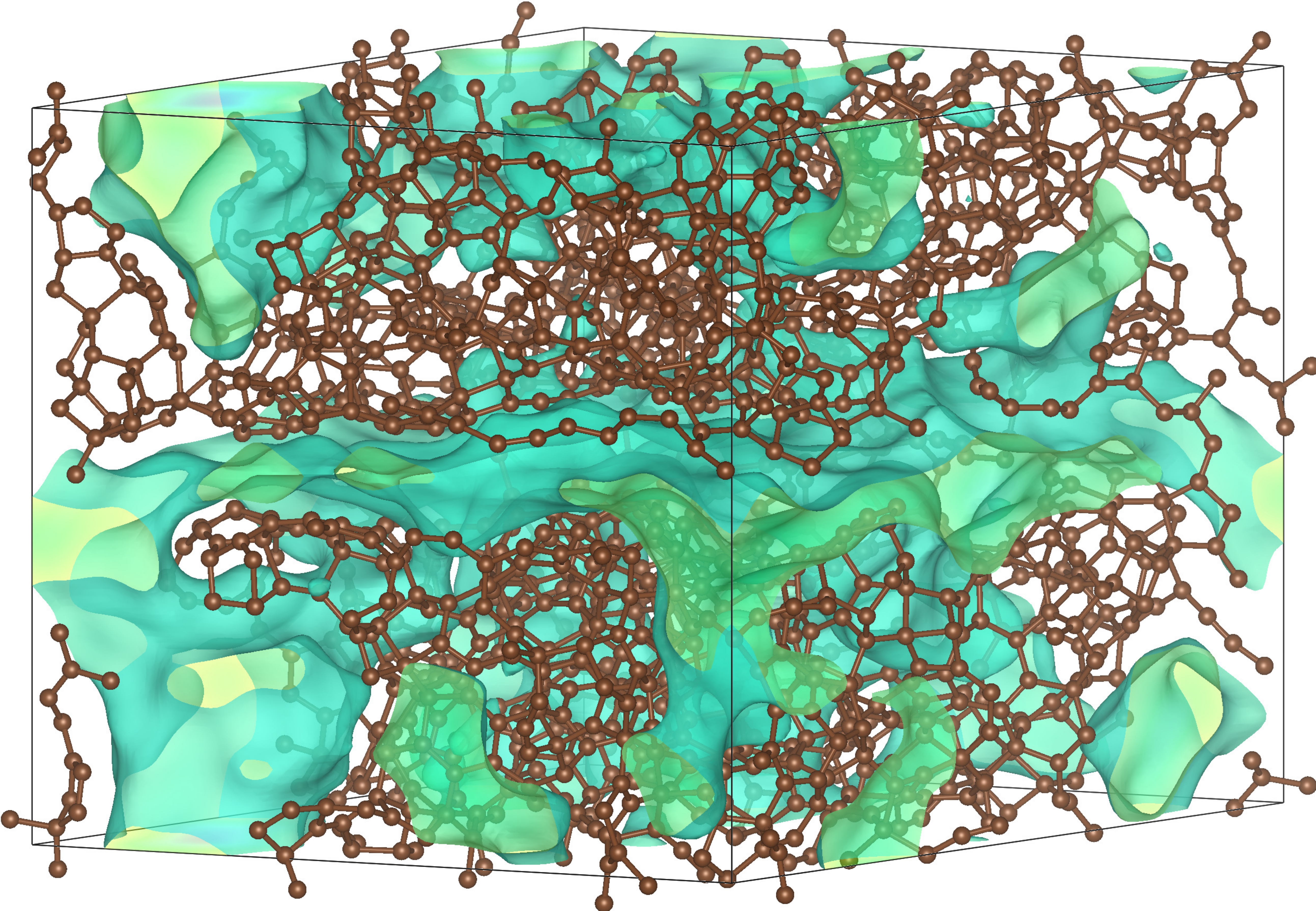}\label{fig:structure:C}}%
    \hspace{1em}%
    \subfloat[Structure E]{\includegraphics[width=0.95\linewidth]{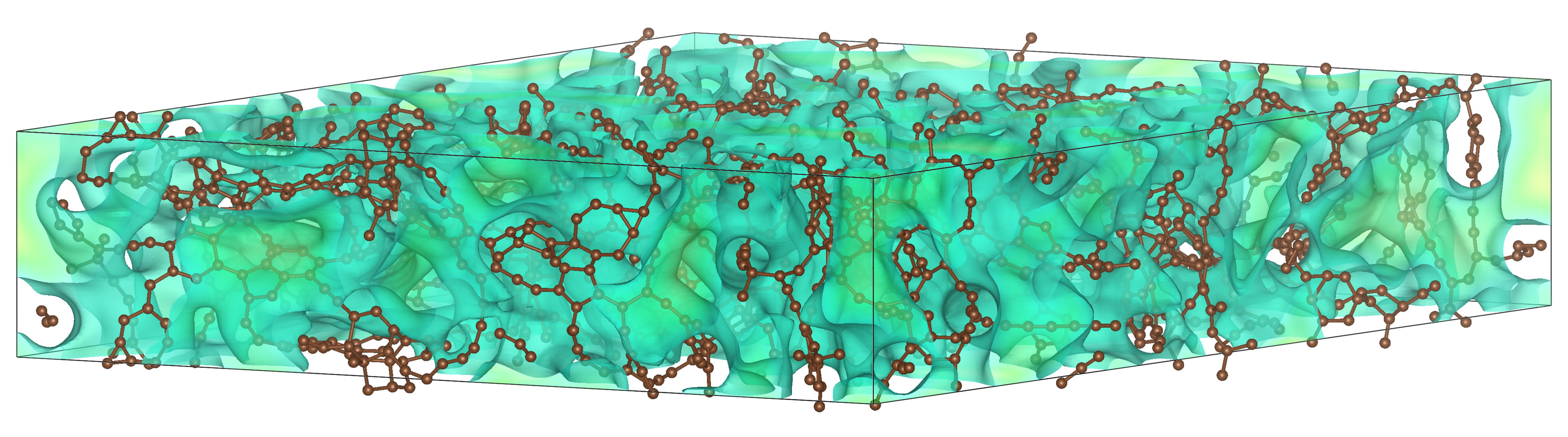}\label{fig:structure:E}}%
    \caption{%
    Ball-and-stick model representation of structures \protect\subref{fig:structure:B} B, \protect\subref{fig:structure:C} C, and \protect\subref{fig:structure:E} E under zero intercalaction with isosurfaces representing the pore network; isosurfaces are rendered for points in space with a minimum atomic distance of roughly 1.8~\si{\angstrom}.%
    }
    \label{fig:structure}
\end{figure*}

The atomic structure and pore network for structures B, C, and E are presented in \figref{fig:structure} (structures A and D are presented in the main article). Structures C and D were generated with a sacrificial nanotube template, producing a larger connected pore. The set spans different densities, porosities, and pore morphologies while retaining experimentally relevant sp$^3$ fractions.

The numerical values for density, sp$^3$ fraction porosity, and formation energy of the five hard carbon structures A--E are provided in \tabref{tab:structure}.

\begin{table}[htbp]
\centering
\caption{Structural and energetic properties of the hard carbon structures A--E.}
\label{tab:structure}
\begin{tabular}{@{} l c c c c @{}}
\toprule
Structure\hspace{1em} & Density (\si{\gram\per\centi\metre\squared}) & sp$^3$ fraction & Porosity (\%) & Formation energy (\si{\electronvolt\per{atom}}) \\
\midrule
A & 2.484 & 0.386 & 1.64 & 0.812 \\
B & 2.031 & 0.300 & 8.04 & 0.870 \\
C & 2.031 & 0.324 & 8.79 & 0.865 \\
D & 1.727 & 0.278 & 15.18 & 0.894 \\
E & 1.246 & 0.107 & 20.72 & 1.046 \\
\bottomrule
\end{tabular}
\end{table}

\section{Porosity}

\begin{figure}
    \centering
    \includegraphics[width=0.5\linewidth]{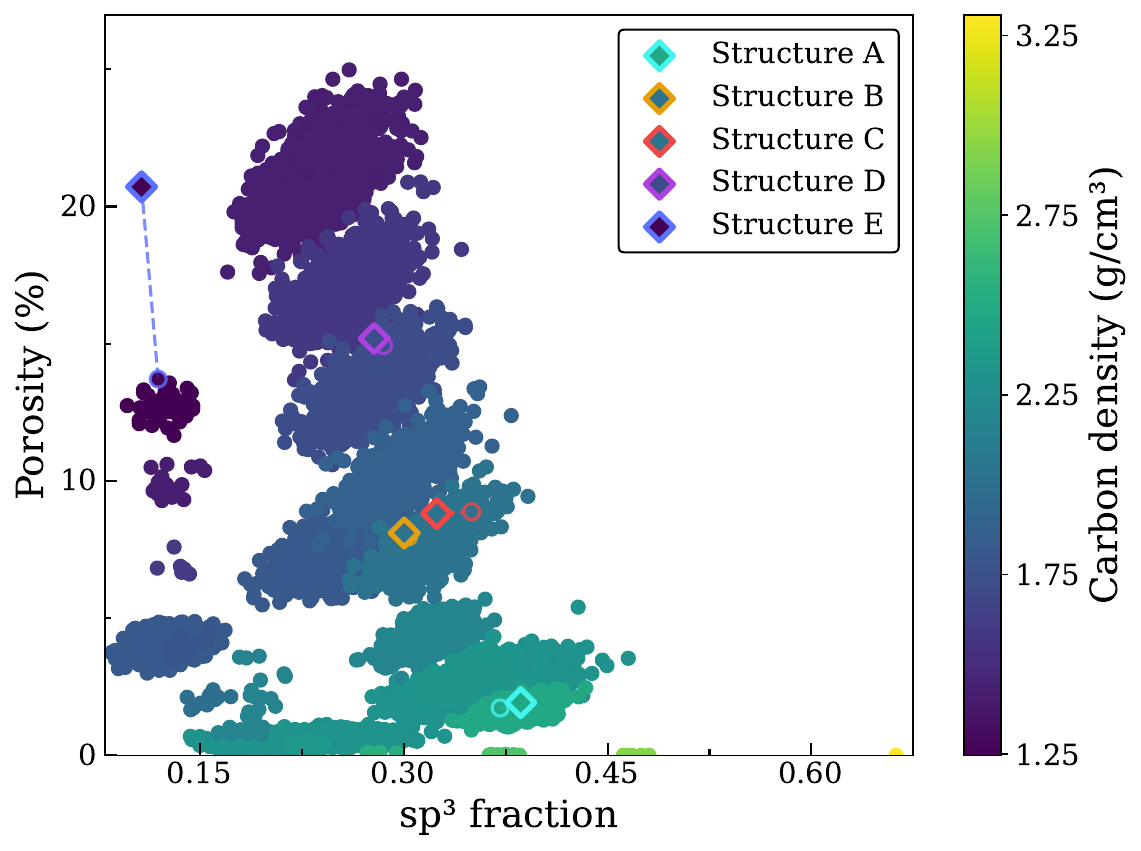}
    \caption{Extended figure of porosity for all generated structures. Equivalent figure in the main paper cuts off the point at porosity = 0 and sp$^3$ fraction of 0.6623 to focus in on structures A-E.}
    \label{fig:porosity:extd}
\end{figure}

To show the one data point (at porosity = 0 and sp$^3$ fraction of 0.6623) we exclude from the main text (Fig. 1b) to focus on structures A--E, we present \figref{fig:porosity:extd}.

\section{Density}

\tabref{tab:sodium_density} shows sodium density statistics for hard carbon structures under each structure's maximum capacity.

\begin{table}[htbp]
\centering
\caption{Sodium density statistics for hard carbon structures under each structure's maximum capacity intercalation limit. Relative standard deviation is given as the standard deviation divided by the mean density.}
\label{tab:sodium_density}
\begin{tabular}{@{} l c c c c @{}}
\toprule
\multirow{2}{*}{Structure}\hspace{1em} & \multirow{2}{*}{\makecell{Mean density \\ (\si{\gram\per\centi\metre\cubed})}} & \multicolumn{2}{c}{Std dev (\si{\gram\per\centi\metre\cubed})} & \multirow{2}{*}{Relative std (\%)} \\
\cmidrule(lr){3-4}
 & & {\hspace{1em}Value\hspace{1em}} & {\hspace{1em}Range\hspace{1em}} & \\
\midrule
A & $0.09 \pm 0.00$ & $0.07 \pm 0.01$ & 0.06--0.08 & $76.1 \pm 10.0$ \\
B & $0.38 \pm 0.00$ & $0.17 \pm 0.06$ & 0.01--0.26 & $45.4 \pm 15.9$ \\
C & $0.41 \pm 0.00$ & $0.19 \pm 0.04$ & 0.14--0.25 & $47.6 \pm 10.7$ \\
D & $0.54 \pm 0.00$ & $0.18 \pm 0.04$ & 0.15--0.24 & $33.3 \pm 7.4$ \\
E & $0.82 \pm 0.00$ & $0.12 \pm 0.04$ & 0.06--0.15 & $14.1 \pm 4.9$ \\
\bottomrule
\end{tabular}
\end{table}

\begin{figure*}
    \centering
    \vspace{-0.9in}
    \hspace*{-0.9in}
    \subfloat[Structure A]{\includegraphics[width=1.25\linewidth]{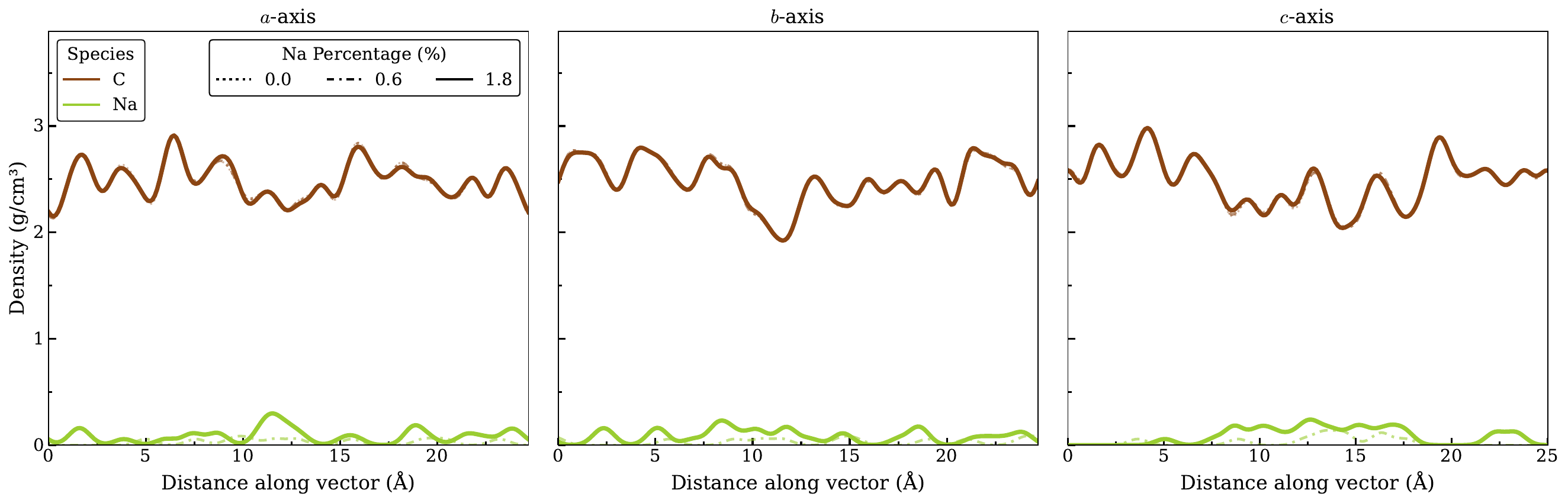}}
    \hfill
    \hspace*{-0.9in}
    \subfloat[Structure B]{\includegraphics[width=1.25\linewidth]{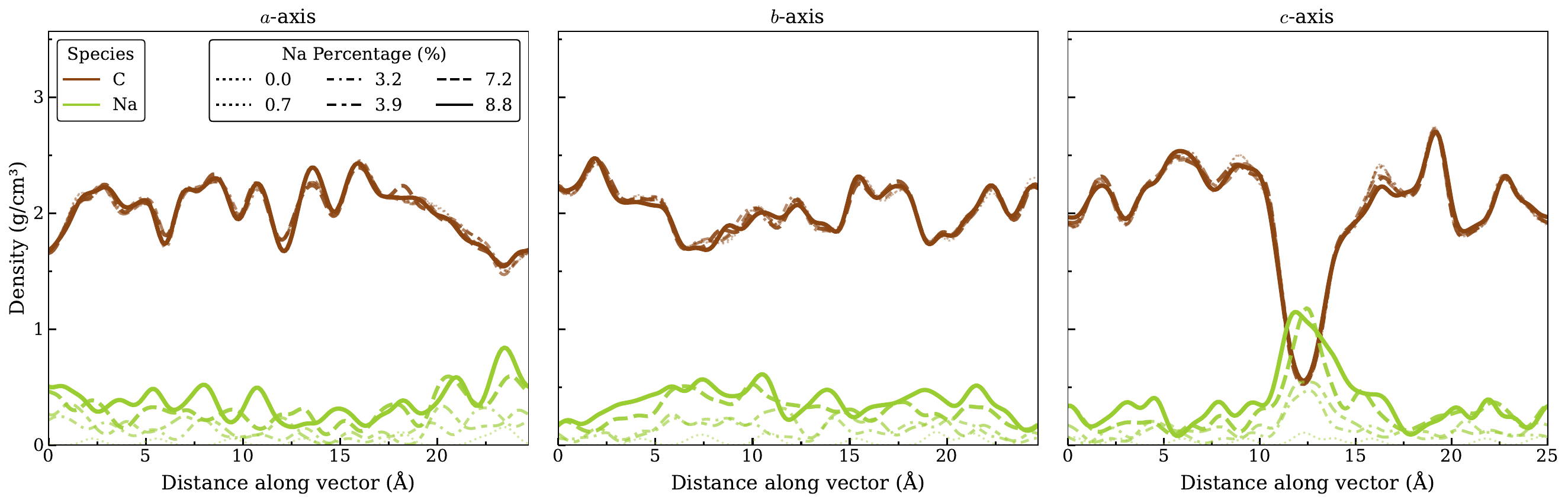}}
    \hfill
    \hspace*{-0.9in}
    \subfloat[Structure C]{\includegraphics[width=1.25\linewidth]{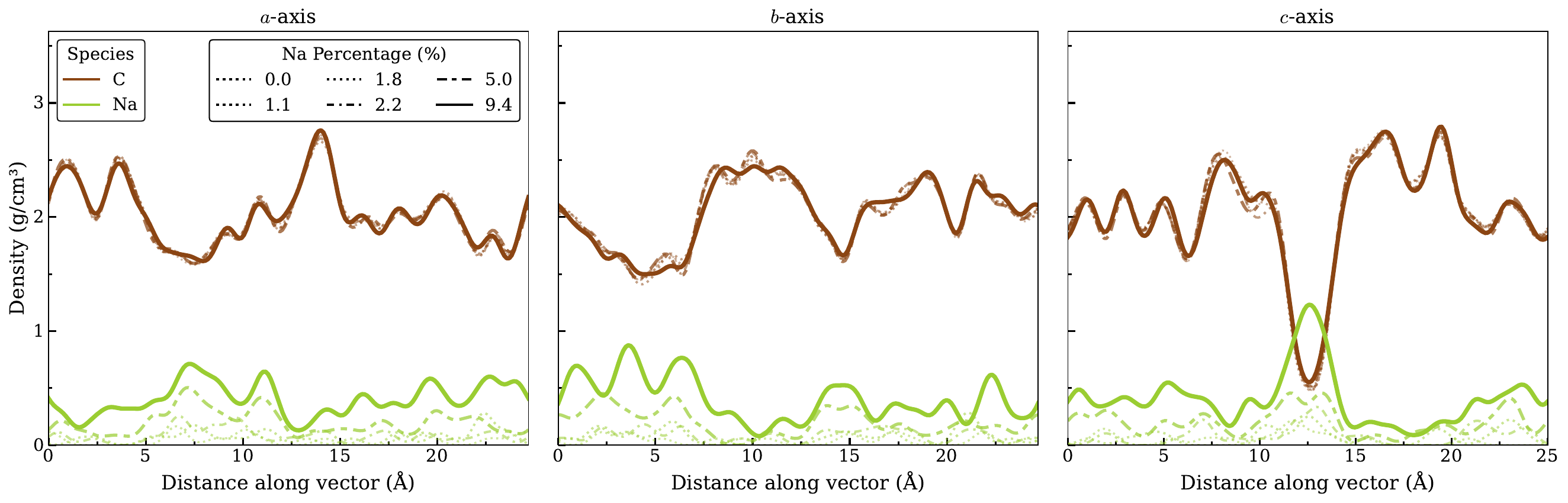}}
    \caption{Shows the planar densities of Carbon and Sodium along the a-, b-, and c-axis for structure A (a), B (b), and C (c).}
    \label{suppfig:planardensity1}
\end{figure*}

\figrefs{suppfig:planardensity1}{suppfig:planardensity2} shows the planar densities of Carbon and Sodium along the a-, b-, and c-axis for structure A--C and structures D \& E, respectively.

\begin{figure*}
    \centering
    \vspace{-0.9in}
    \hspace*{-0.9in}
    \subfloat[Structure D]{\includegraphics[width=1.25\linewidth]{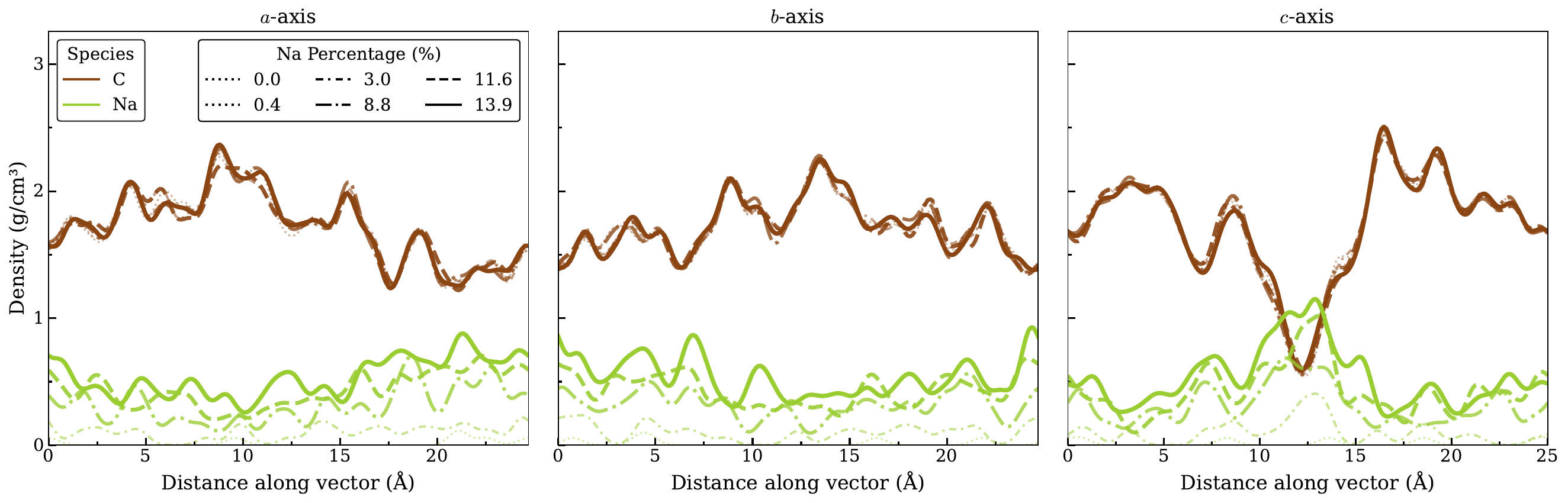}}
    \hfill
    \hspace*{-0.9in}
    \subfloat[Structure E]{\includegraphics[width=1.25\linewidth]{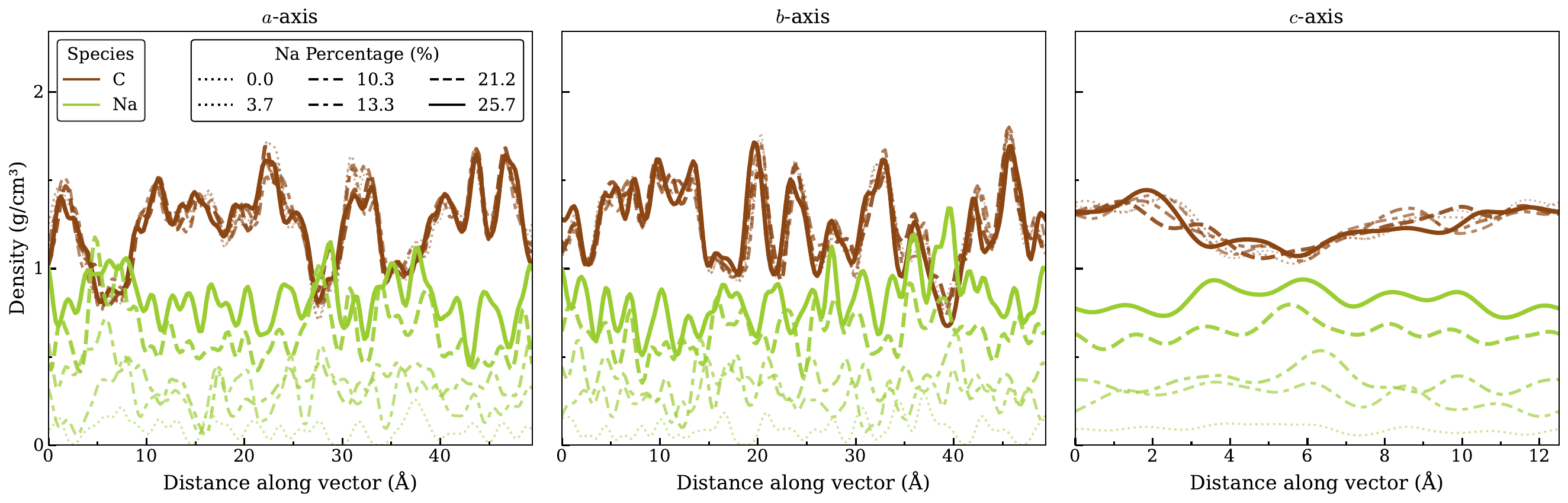}}
    \caption{Shows the planar densities of Carbon and Sodium along the a-, b-, and c-axis for structure D (a), and E (b).}
    \label{suppfig:planardensity2}
\end{figure*}

\section{Pore shape}

\begin{figure*}
    \centering
    \subfloat[]{\includegraphics[width=0.75\linewidth]{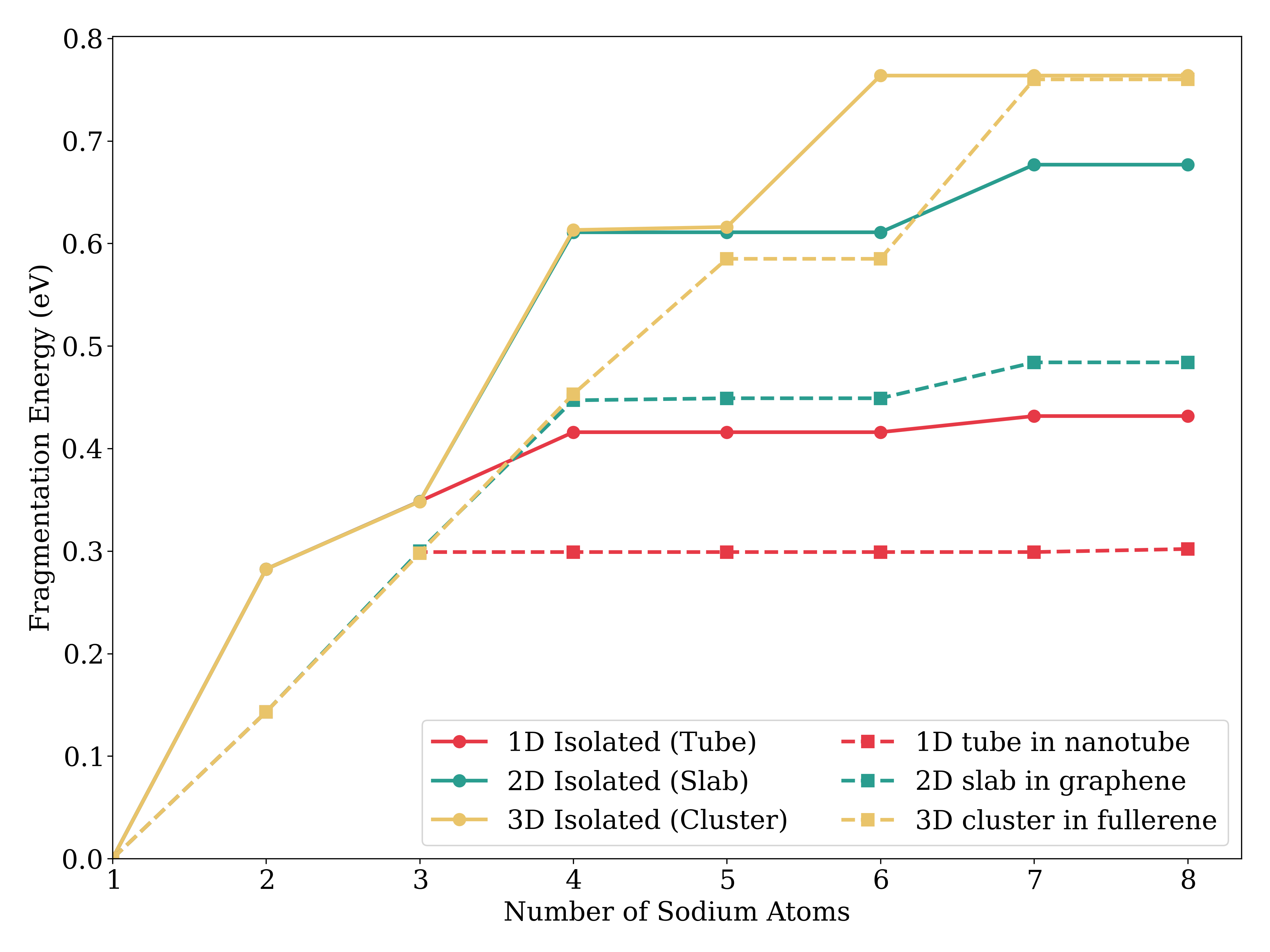}\label{fragmentation_energy_bare_vs_carbon}}
    \hfill
    \subfloat[]{\includegraphics[width=0.75\linewidth]{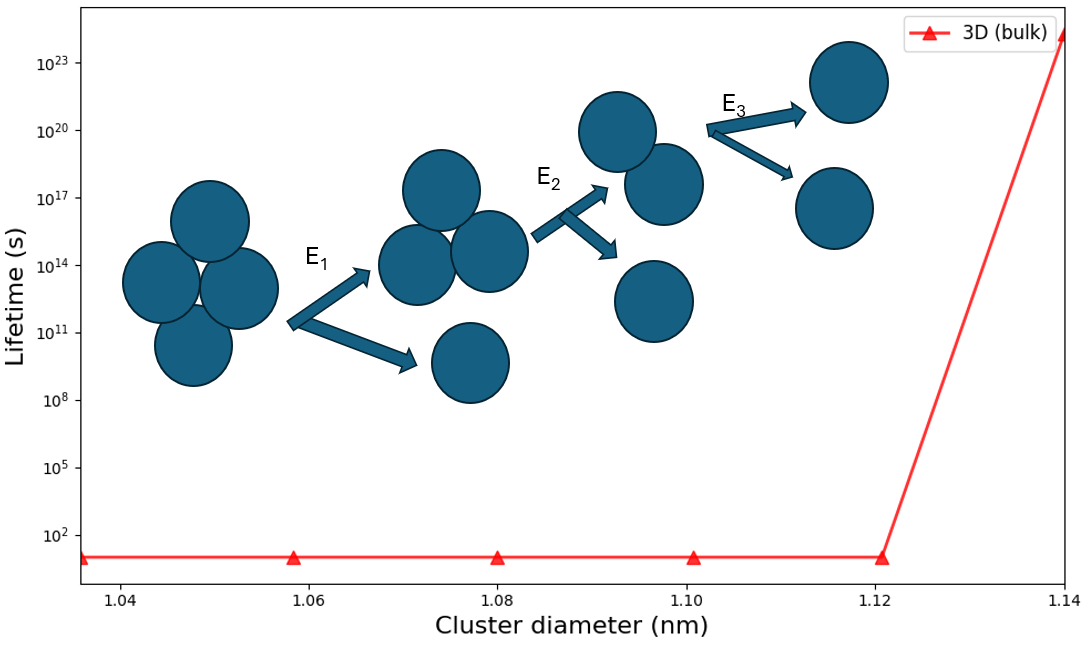}\label{Lifetimewithexample}}
    \caption{Effect of pore geometry on sodium-cluster reversibility. (a) Fragmentation energy as a function of cluster size for RAFFLE-generated sodium clusters under 1D tubular, 2D slab, and 3D compact constraints, both isolated (dotted lines) and confined by carbon (solid lines). (b) Cluster lifetime on a logarithmic scale as a function of cluster diameter for the 1D chain geometry, with schematic illustrations of the stepwise fragmentation pathway.}
    \label{fig:frag_lifetime}
\end{figure*}

Pore geometry strongly influences sodium-cluster stability and reversibility. Fragmentation energy, which measures resistance to dissolution, increases monotonically with cluster size across all configurations (\figref{fig:frag_lifetime}), reflecting the larger number of Na--Na bonds and the increasing metallic character of larger clusters. For clusters templated from metallic sodium, the cluster lifetime rises sharply at a size corresponding to a pore diameter of approximately 1.1~nm, comparable to pore sizes associated with efficient plateau storage in hard carbons~\cite{Bommier2014,Gan2025}. The metallic template is less reliable for small or strongly non-spherical clusters, but is appropriate for larger near-spherical pores.

A clear geometry hierarchy emerges: compact 3D clusters have the highest fragmentation energies, followed by 2D slab-like clusters, while elongated 1D tubular clusters are least stable. At six sodium atoms, typical fragmentation energies are approximately 0.5~eV for 3D clusters, 0.35~eV for 2D clusters, and 0.25~eV for 1D clusters. Higher fragmentation energies imply larger barriers to cluster dissolution during desodiation, promoting irreversible sodium trapping and reducing both reversible capacity and initial Coulombic efficiency. The $3\mathrm{D} > 2\mathrm{D} > 1\mathrm{D}$ ordering therefore suggests that hard carbons dominated by elongated pore geometries should show lower irreversible nucleation and improved reversibility. Carbon confinement increases the absolute fragmentation energies across all geometries but preserves the same hierarchy (\figref{fig:frag_lifetime}), indicating that pore shape remains a useful design descriptor under more realistic conditions. This result is consistent with the broader view that nanoscale carbon morphology controls alkali-ion storage~\cite{Bruce2008,DouHardCarbonReview2019,Gan2025} and suggests that pore-shape control can improve sodium-ion anodes.

\section{Sodium-storage mechanism in a model hard carbon}
\begin{figure*}
    \centering
    \subfloat[]{\includegraphics[width=0.45\linewidth]{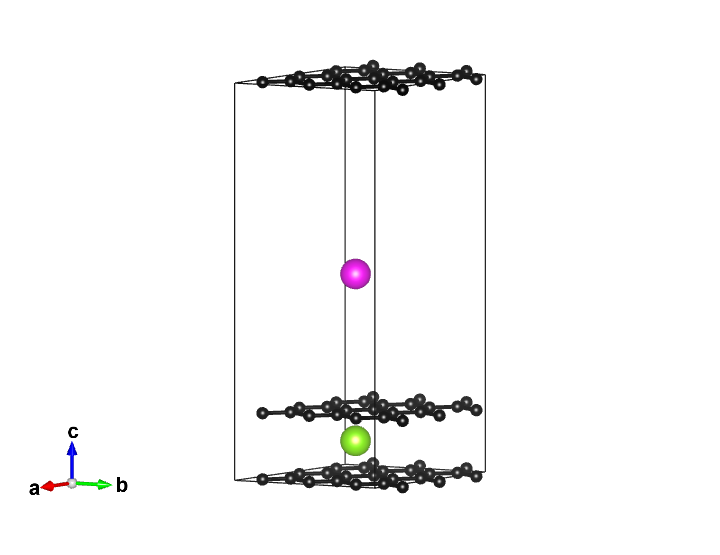}\label{subfig:Intercalationpore}}
    \subfloat[]{\includegraphics[width=0.45\linewidth]{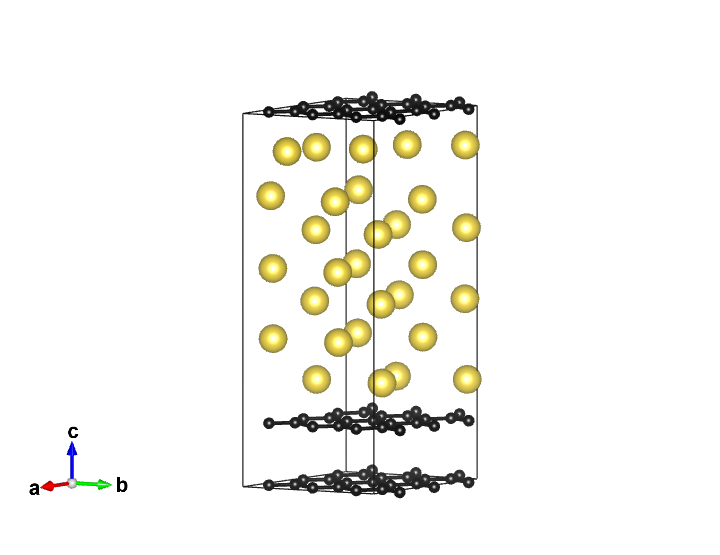}\label{subfig:relaxed_Na29}}
    \caption{Model pore-interlayer geometry used to probe sodium storage. (a) A Na$^+$ ion in the pore region and a Na$^+$ ion in the graphitic interlayer. (b) Pore-saturated configuration after relaxation, showing the structural response of the graphitic spacing.}
    \label{fig:vesta_combined}
\end{figure*}

We next examined sodium-storage mechanisms using DFT calculations on a bilayer hard carbon motif containing both a pore and a graphitic interlayer (\figref{fig:vesta_combined}). Formation energies were computed for Na$^+$ occupying each environment. In the unsaturated structure, sodium in the pore gives $E_f = -1.076$~eV, indicating strong stabilisation relative to metallic sodium, while interlayer sodium gives $E_f = -0.117$~eV. After pore saturation, the interlayer formation energy increases to $E_f = +1.163$~eV, a shift of approximately 1.28~eV that makes intercalation thermodynamically unfavourable.

This change arises from coupling between pore filling and interlayer accessibility. Relative to the 3.35~\AA{} spacing characteristic of graphite~\cite{HarrisPerspectivesGraphiticCarbon2005}, insertion of a single Na$^+$ in the unsaturated model expands the local interlayer spacing to 5.5773~\AA{}. Once the pore is saturated, relaxation contracts the spacing to 4.2371~\AA{}, blocking further intercalation. The model therefore predicts that intercalation is suppressed after pore filling, consistent with experimental and review evidence that the low-voltage plateau is dominated by pore-related storage rather than extensive graphitic staging~\cite{SaurelChargeStorageMechanism2018,Gan2025,Tan2023}; sustained intercalation would be expected to produce a measurable expansion. Spectroscopic studies of hard carbon sodiation also indicate limited restructuring of the graphitic network during plateau storage~\cite{LiHeteroatomDoping2017,Gan2025}. Models that treat intercalation and pore filling as independent parallel processes~\cite{SaurelChargeStorageMechanism2018,Gan2025} may therefore overestimate the intercalation contribution. Independent intercalation could still occur in large, isolated graphitic domains, but the reversible capacity of sodium in graphite is only approximately 35~mAh~g$^{-1}$~\cite{Yao2024}, so pore filling remains the dominant storage contribution in high-capacity hard carbons.

\section{Minimising computation to maximise workflow}
\label{Sec:datasetbench}
The A--D structures discussed in the main text were fully relaxed at every stage of sodium intercalation, as described above. While accurate, this procedure is too computationally expensive for generating the thousands of data points needed to train a neural network.

To reduce cost, we tested how many relaxation steps were required per intercalation stage to reproduce the voltage and capacity trends of the fully relaxed calculations. Three scenarios were compared: no relaxation, 10 relaxation steps, and 50 relaxation steps.

The procedure was as follows. For each host structure, all void sites large enough to accommodate a sodium ion were identified. The maximum possible sodium concentration was estimated from the total void volume, and the range from zero to this maximum was divided into 10 equal composition steps. At each step, 10 random configurations were generated by placing sodium atoms at distinct void sites. These configurations were then subjected to either no relaxation, 10 relaxation steps, or 50 relaxation steps. The lowest energy among the 10 configurations at each step was used to calculate the voltage, and the standard deviation of the 10 energies provided an estimate of the uncertainty.

\begin{figure*}
    \centering
    \includegraphics[width=\linewidth]{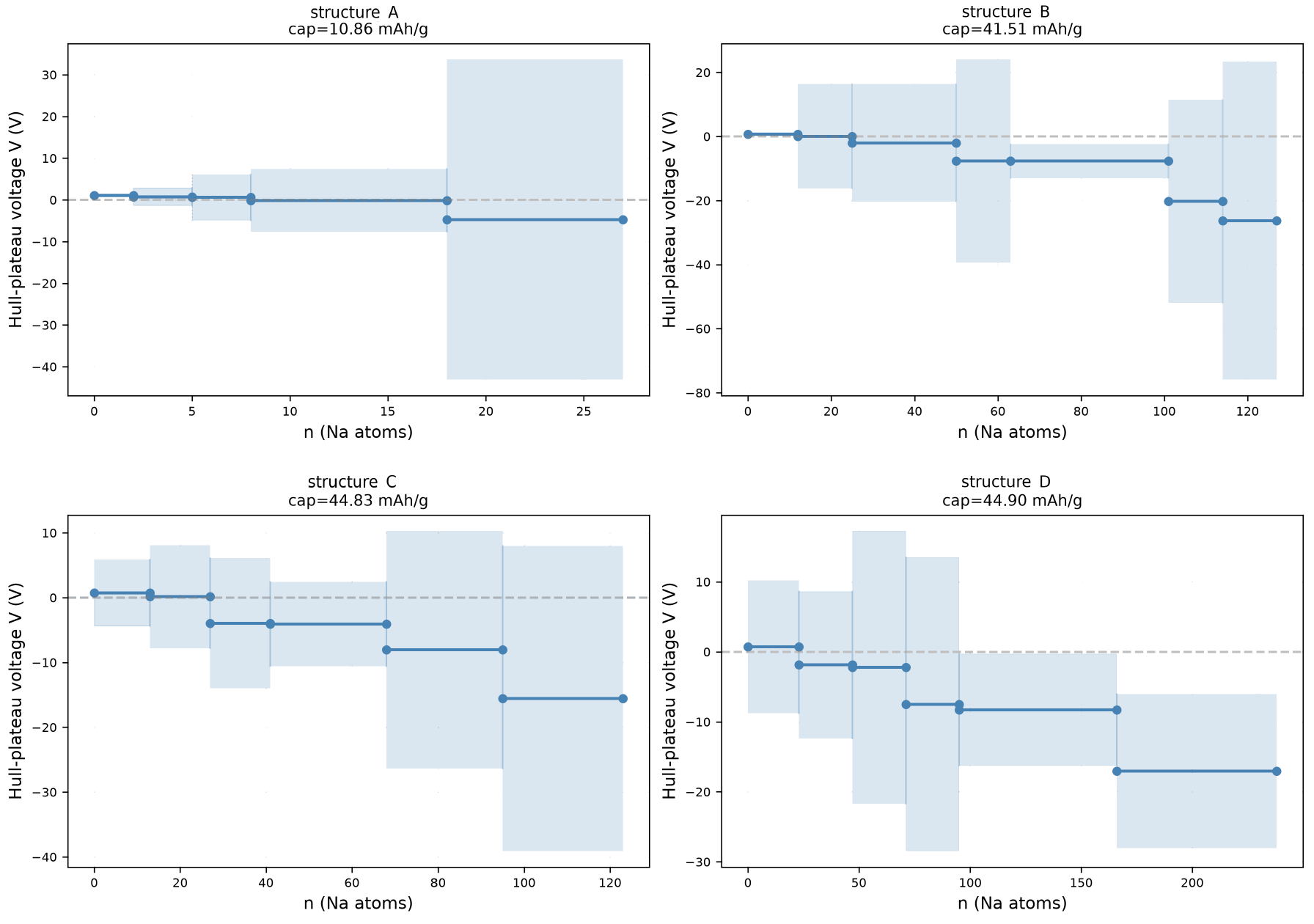}
    \hfill
    \caption{Voltage versus capacity for structures A--D calculated with no relaxation.}
    \label{suppfig:A-DNoRelax}
\end{figure*}

\begin{figure*}
    \centering
    \includegraphics[width=\linewidth]{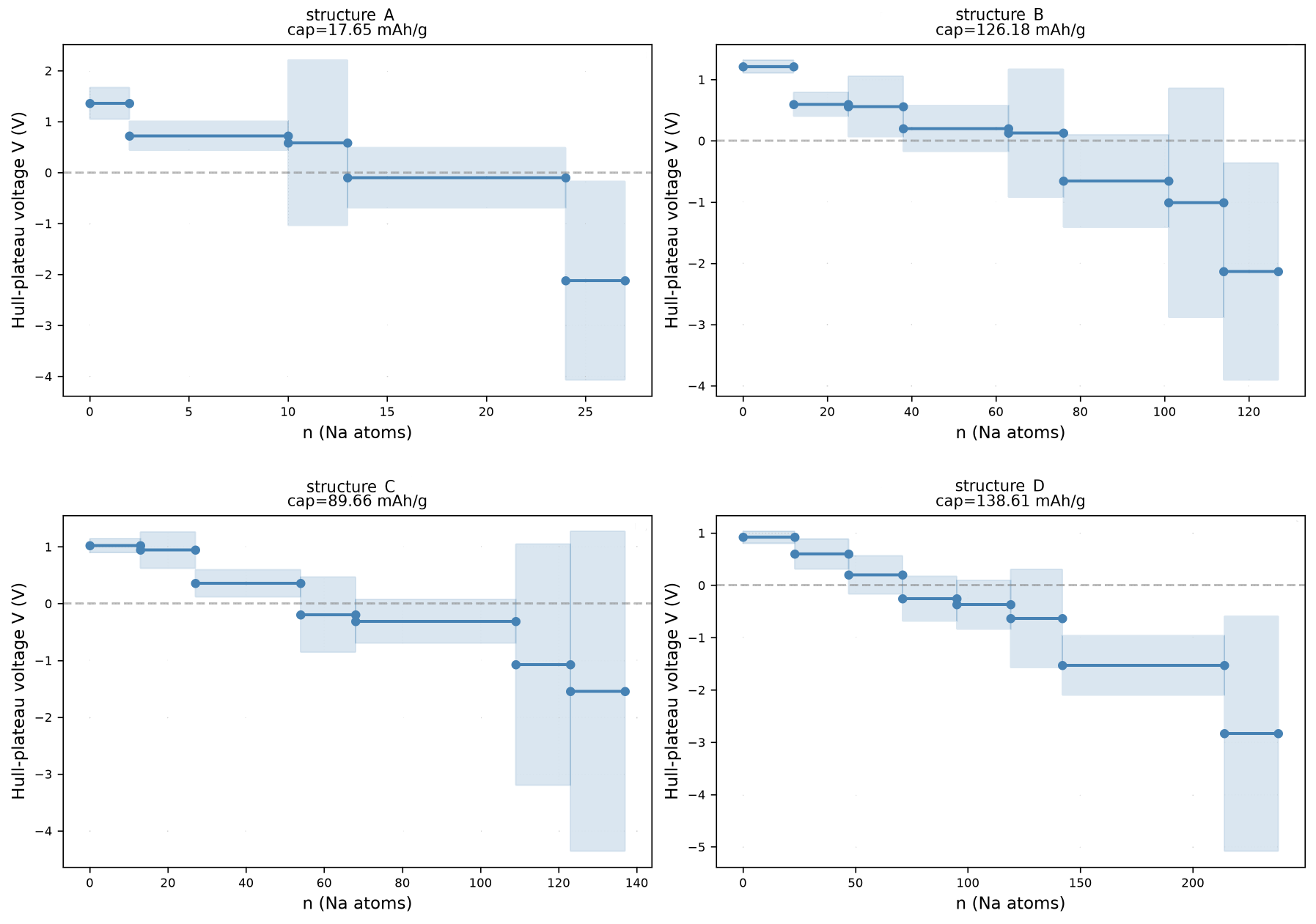}
    \caption{Voltage versus capacity for structures A--D calculated with 10 relaxation steps.}
    \label{suppfig:A-D10Relax}
\end{figure*}

\begin{figure*}
    \centering
    \includegraphics[width=\linewidth]{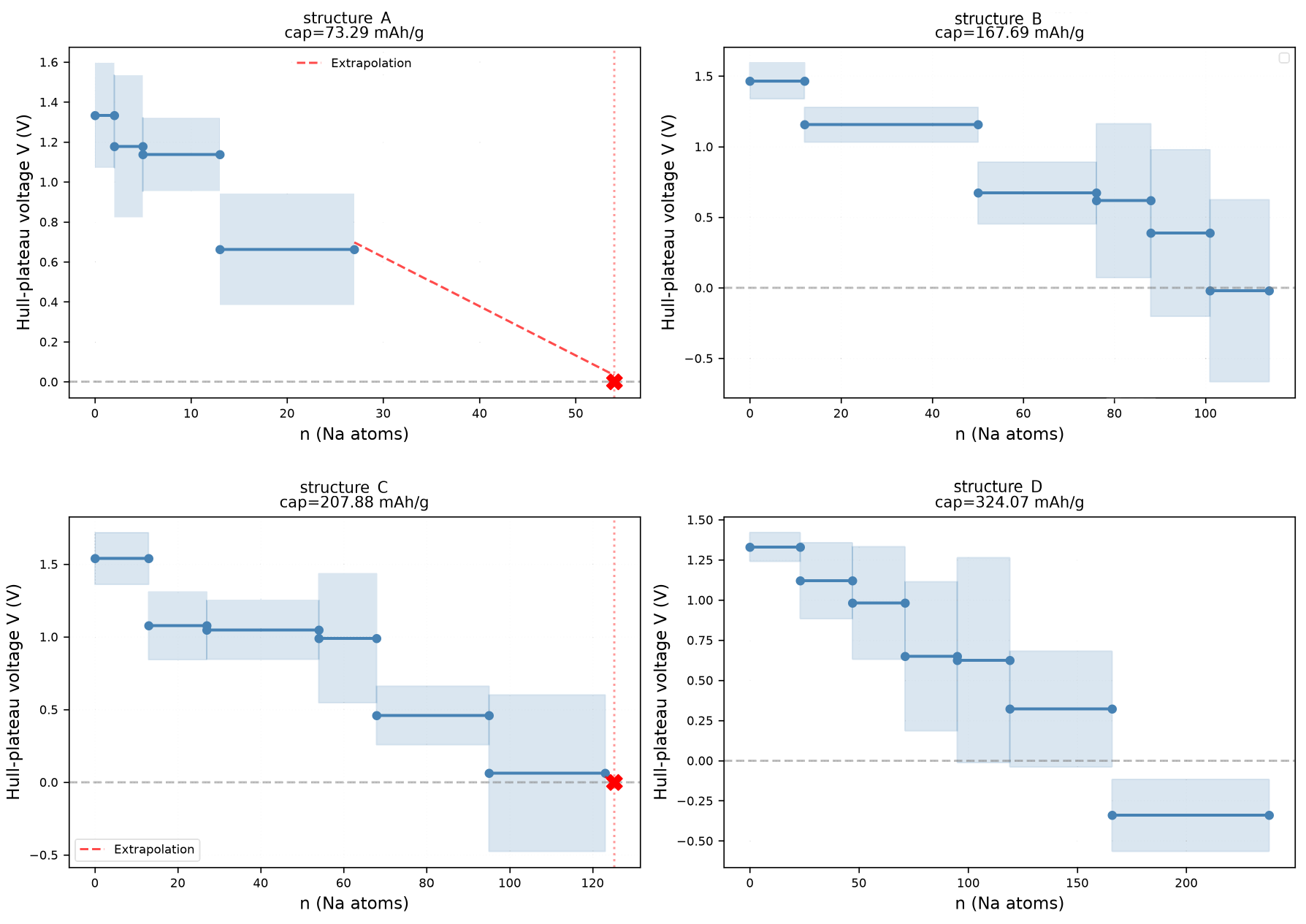}
    \caption{Voltage versus capacity for structures A--D calculated with 50 relaxation steps.}
    \label{suppfig:AD50Relax}
\end{figure*}

As shown in \figref{suppfig:A-DNoRelax} and \figref{suppfig:A-D10Relax}, both zero and 10 relaxation steps failed to reproduce the voltages and capacities obtained from the fully relaxed calculations. However, 50 relaxation steps (\figref{suppfig:AD50Relax}) were sufficient to approach the fully relaxed results and to recover the correct ordering of capacities from highest to lowest. This reduced relaxation protocol was therefore adopted to build a database of voltages and capacities for 64 systems, which served as the training set for the neural network.

\section{Methodology: AI model for capacity}
\label{sec:AImethods}

\subsection{Overview}

The computational workflow proceeds in two stages. First, a small but diverse set of porous carbon host structures is labelled with thermodynamic sodium capacities obtained from explicit intercalation calculations using a machine-learned interatomic potential (MLiP). An MLiP is a fast, artificial intelligence-based surrogate for quantum mechanical energy calculations: it is trained on accurate first-principles data but can evaluate the energy of an atomic structure in a fraction of the time, making large-scale screening feasible. Second, a lightweight statistical model is trained to predict capacity directly from the empty host structure, enabling high-throughput screening of vast structural libraries without repeating the expensive intercalation calculations.

\subsection{Thermodynamic sodium capacity}
\label{sec:thermo-capacity}

\subsubsection{Void detection and Na placement}

For each empty carbon host, a fine grid of points (spaced $0.2\;\text{\AA}$ apart) is constructed across the simulation cell. A grid point is classified as a void if it lies farther than $2.18\;\text{\AA}$ from every host atom, ensuring the sodium ion does not overlap with the carbon framework. These void points are grouped into distinct pockets using a clustering algorithm (DBSCAN, with $\varepsilon = 3.0\;\text{\AA}$ and a minimum of 3 points per pocket), and any pocket smaller than $8\;\text{\AA}^3$ is discarded as too small to host an ion.

The maximum number of sodium atoms that can be inserted is estimated from the total void volume $V_{\text{void}}$ and an empirical atomic volume for sodium of $v_{\text{Na}} = 10\;\text{\AA}^3$:
\begin{equation}
  n_{\max} = \min\!\left(\Bigl\lfloor\frac{V_{\text{void}}}{v_{\text{Na}}}\Bigr\rfloor,\; N_{\text{void points}}\right).
\end{equation}
The composition range $n = 1, \dots, n_{\max}$ is sampled in $n_{\text{steps}} = 10$ steps. At each step, up to 10 candidate structures are generated by placing $n$ sodium atoms at distinct random void points. Each candidate is evaluated with the MACE-MP-0 medium model, a general-purpose MLiP, supplemented by D3 dispersion corrections to account for long-range van der Waals interactions. Optionally, each candidate is relaxed with the FIRE2 algorithm (maximum step $0.2\;\text{\AA}$, force tolerance $0.02\;\text{eV\,\AA}^{-1}$, 50 steps). The lowest intercalation energy $E(n)$ and the mean and standard deviation over the 10 candidates are recorded.

\subsubsection{Convex-hull construction and specific capacity}

To determine the thermodynamic stability of each sodium concentration, formation energies relative to bulk metallic sodium are computed as
\begin{equation}
  E_f(n) = E(n) - E_{\text{host}} - n\,E_{\text{Na}}^{\text{bulk}},
  \label{eq:formation-energy}
\end{equation}
with $E_f(0) \equiv 0$. The lower convex hull of the set $\{(n, E_f(n))\}$ is then constructed. The convex hull is a geometric tool that identifies the most stable compositions: any point lying on the hull represents a composition that is more stable than any linear combination of neighbouring compositions.

The open-circuit plateau voltage between adjacent hull vertices $i \to j$ is given by the slope of the hull segment:
\begin{equation}
  V_{i\to j} = -\frac{E_f(n_j) - E_f(n_i)}{n_j - n_i}.
\end{equation}
The capacity composition $n_{\text{capacity}}$ is defined as the largest hull vertex whose preceding segment still exhibits a positive voltage ($V > 0$). If the final hull segment remains positive, a linear extrapolation is performed by fitting the last three plateau voltages and solving for $V = 0$, capped at twice the last sampled $n$ to yield $n_{\text{capacity}}^{\text{extrap}}$. The specific capacity (per gram of host carbon) is then
\begin{equation}
  C = \frac{n_{\text{capacity}}^{\text{extrap}} \times 26\,801.6}{M_{\text{host}}}
  \quad [\text{mAh g}^{-1}],
\end{equation}
where $M_{\text{host}}$ is the molar mass of the carbon host (\si{\gram\per\mol}), and the numerical prefactor converts electrons per formula unit to the standard battery unit of ~\si{\milli\ampere\hour\per\gram}. Unphysical data (energies exceeding $10^{6}$~\si{\electronvolt}, or numerical NaN/Inf values) are rejected before hull analysis.

\subsection{MACE--GNN surrogate model}
\label{sec:surrogate}

\subsubsection{Feature generation}

The surrogate model must learn to predict capacity from the structure of the empty carbon host alone, without performing any sodium intercalation calculations. To achieve this, each host structure is first converted into a numerical fingerprint by a frozen MACE-MP-0 small model. MACE (Message Passing Atomic Cluster Expansion) is a type of neural network that computes a vector descriptor for each atom based on its local chemical environment. These per-atom descriptors are pooled across the entire structure by taking the mean, maximum, and standard deviation over all atoms, yielding a 768-dimensional vector that captures the global structural character of the host.

Because 768 dimensions is large relative to the size of the training set (64 structures), principal component analysis (PCA) is used to compress the fingerprint to 24 components while retaining more than 99\% of the original information. PCA is a standard statistical technique that identifies the directions of greatest variation in the data and projects the descriptors onto these directions, discarding redundant or noisy dimensions.

Five simple geometric descriptors are then appended to the compressed fingerprint: the total number of void points, the number of distinct void pockets, the total void volume, a heuristic capacity estimate (void volume divided by $10\;\text{\AA}^3$ per sodium), and the simulation cell volume. The combined feature vector is standardised (shifted to zero mean and unit variance) using training-set statistics only, ensuring the model receives inputs on a consistent scale.

\subsubsection{Architecture}

The surrogate is a compact residual multilayer perceptron (MLP), which is a type of feedforward neural network. It includes a learnable linear connection from the standardised heuristic capacity estimate:
\begin{equation}
  \hat{y} = \alpha\,\tilde{C}_{\text{est}} + \beta + \delta\!\bigl(h(\mathbf{x})\bigr),
\end{equation}
where $h(\mathbf{x})$ is a two-layer network with 48 and 24 hidden units, using the GELU activation function, layer normalisation (which stabilises training by keeping layer inputs well scaled), dropout regularisation (randomly zeroing 35\% of neurons during training to prevent overfitting), and an identity-mapped residual connection that helps gradients flow through the network. The term $\alpha\,\tilde{C}_{\text{est}} + \beta$ acts as a physical prior, anchoring the prediction to the simple geometric estimate when data are scarce, while the neural network $\delta(h(\mathbf{x}))$ learns corrections based on the full structural fingerprint.

The network has three output heads: a capacity mean, a log-variance (clamped to the range $[-6, 4]$ for numerical stability), and a capacity standard deviation obtained via a softplus activation (which ensures positive values). The total training loss combines three objectives:
\begin{equation}
  \mathcal{L} = 0.5\,\mathcal{L}_{\text{NLL}} + 0.5\,\mathcal{L}_{\text{Huber}} + 0.35\,\mathcal{L}_{\text{MSE}}^{\text{(std)}}.
\end{equation}
The negative log-likelihood (NLL) term teaches the model to predict both the mean and its uncertainty; the Huber loss provides robust mean regression; and the mean squared error (MSE) term trains the standard deviation head. Targets are normalised to zero mean and unit variance in the original mAh g$^{-1}$ space so that the heuristic capacity skip remains physically interpretable.

\subsubsection{Training protocol}

Models are trained with the AdamW optimiser (initial learning rate $8\times10^{-4}$, weight decay $5\times10^{-3}$) and gradient clipping (limiting the $\ell_2$ norm of gradients to 1.0) to prevent unstable updates. The learning rate is automatically halved whenever the validation loss stops improving for 12 consecutive epochs, down to a minimum of $10^{-5}$. Early stopping halts training if no improvement is seen for 40 epochs, preventing overfitting.

Model performance is assessed with stratified $k$-fold cross-validation: the data are divided into folds such that each fold contains a representative range of capacities. Out-of-distribution robustness is tested with GroupKFold by structure family (trajectory file), which measures whether the model generalises to entirely new types of carbon hosts. The final deployment model is trained on the majority of the data with a small stratified validation holdout.

\subsection{High-throughput screening}

For screening large structural libraries, each candidate frame is processed by the void detector (or cached void features are reused), encoded by the frozen MACE model, and fed through the trained surrogate. The model outputs a predicted capacity mean, an aleatoric uncertainty (inherent noise in the prediction), and an epistemic uncertainty estimate (uncertainty due to limited training data). Predictions are accumulated with resume support to guard against interruption.

\subsection{Assumptions and limitations}

The reported capacities are equilibrium estimates at 0\,K based on the convex hull construction. Kinetic barriers to sodium diffusion, solid-electrolyte interphase formation, and finite-rate effects are neglected. Extrapolation beyond the sampled composition range assumes a linear voltage decay and is capped at twice the last sampled composition. The surrogate is trained on approximately 60 labelled frames spanning a handful of host families; consequently, out-of-distribution errors (measured by GroupKFold across families) are larger than in-distribution errors. Absolute energies inherit the known biases of the MACE-MP universal potential, but the capacity derivation depends only on relative formation energies [Eq.~\eqref{eq:formation-energy}], which are considerably more robust.

\subsection{Performance}

\begin{figure*}
    \centering
    \subfloat[]{\includegraphics[width=0.48\linewidth]{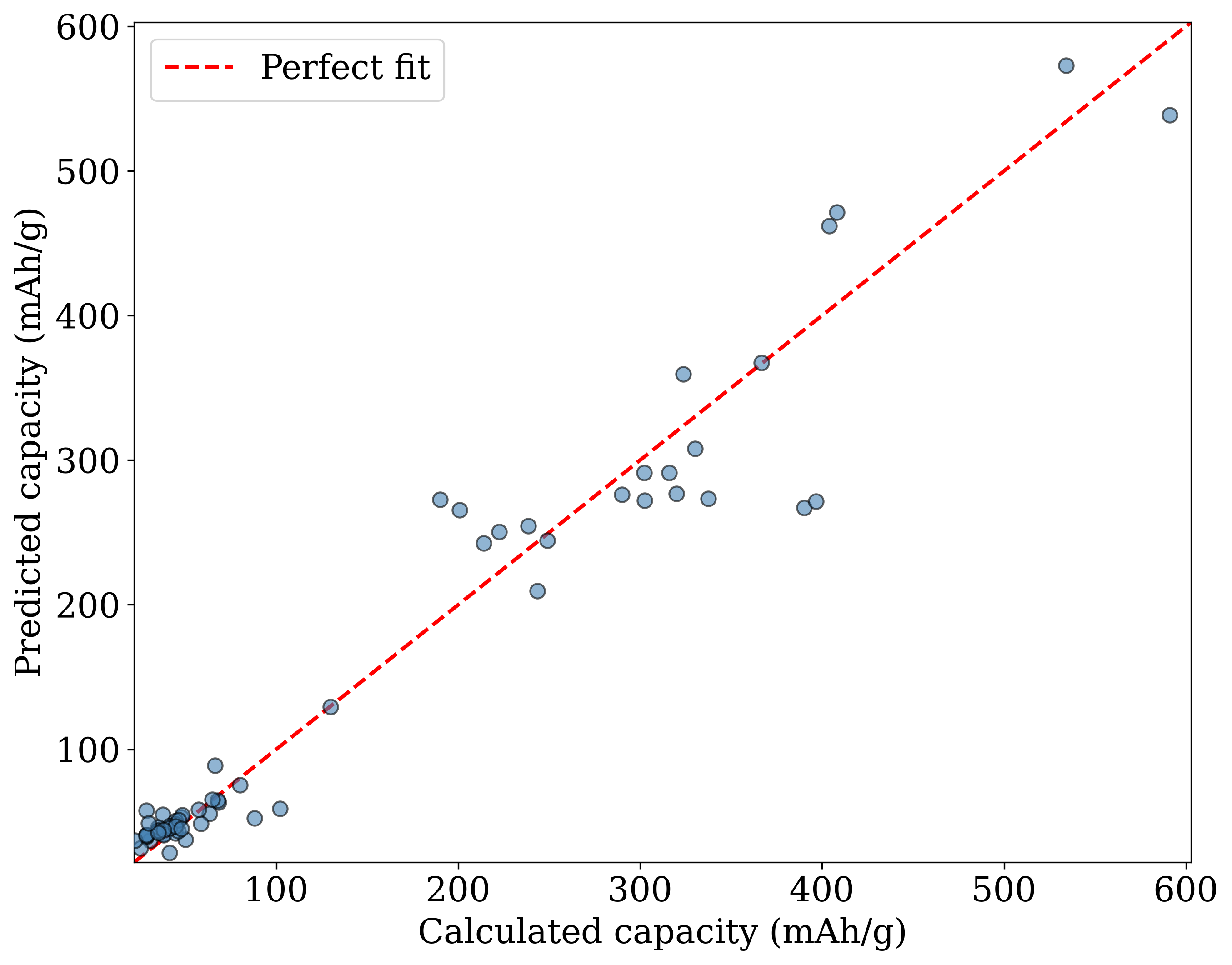}\label{suppfig:capAI}}%
    \hfill
    \subfloat[]{\includegraphics[width=0.48\linewidth]{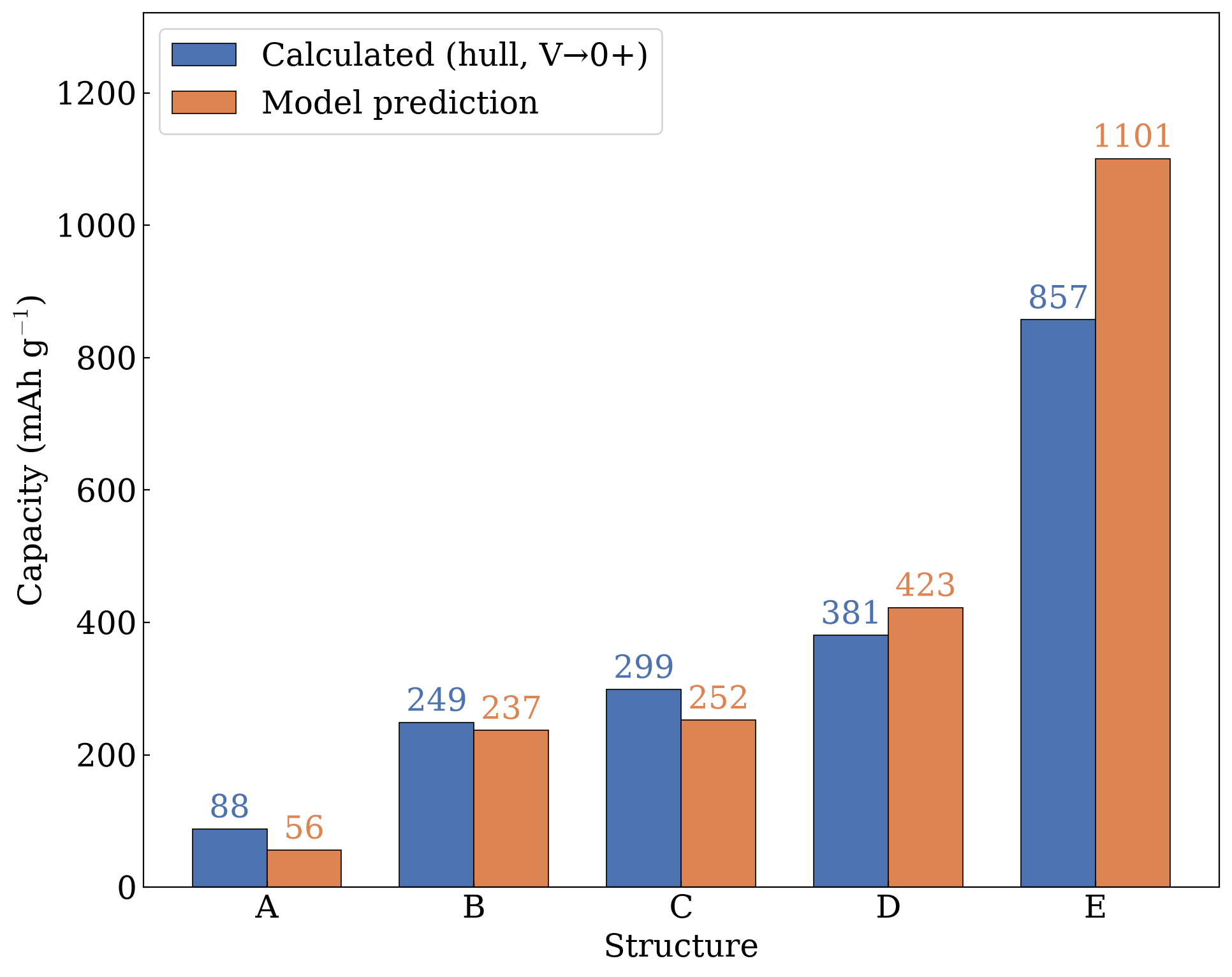}\label{suppfig:AECapModel}}
    \caption{(a) Predicted capacity of the trained neural network compared to the calculated values for the 64 structures. (b) True vs.\ predicted capacity for structures A--E.}
    \label{suppfig:AI}
\end{figure*}

After training on 64 systems, the mean absolute error (MAE) for capacity (\subfigref{suppfig:capAI}) was 21.3\,mAh\,g$^{-1}$, and the coefficient of determination $R^2$ was 0.946. The comparisons between predicted and true capacity for selected structures are shown in \subfigref{suppfig:AECapModel}:

\begin{center}
\begin{tabular}{@{}lccc@{}}
\toprule
& True & Predicted & $\Delta$ \\
\midrule
A & 88.2   & 56.2   & $-32.0$ \\
B & 249.0  & 236.9  & $-12.1$ \\
C & 298.8  & 252.4  & $-46.4$ \\
D & 380.7  & 422.6  & $+41.9$ \\
E & 857.4  & 1100.9  & $+243.5$ \\
\bottomrule
\end{tabular}
\end{center}

\noindent Although the quantitative error is appreciable for some structures, the model retains qualitative predictive power, correctly ranking structures by capacity and identifying high-capacity candidates.

\subsection{Implementation}

The training pipeline is implemented in Python using PyTorch for the neural network, scikit-learn for preprocessing and cross-validation, and the MACE library for descriptor generation. The code automates the following sequence: (1) loading structural data and cached void geometries from a CSV file; (2) encoding each carbon host with the frozen MACE-MP-0 small model to obtain 768-dimensional pooled descriptors; (3) optionally applying PCA dimensionality reduction and appending geometric void features; (4) standardising the combined feature vector; (5) training the residual MLP with the composite loss function, learning rate scheduling, and early stopping; and (6) evaluating performance via stratified and group-based cross-validation before saving the final deployment checkpoint. The saved model can then be loaded to screen new structures by encoding them with the same frozen MACE model, applying the fitted PCA and standardisation, and running a forward pass through the neural network to obtain a capacity prediction in seconds rather than hours/days.

\bibliographystyle{unsrt}
\bibliography{references}